\documentclass[aps,prb,reprint,groupedaddress]{revtex4-2}
\usepackage{amsmath,amssymb,mathtools,bm}
\usepackage{graphicx}
\graphicspath{{./Figures/}}
\usepackage{booktabs,tabularx,array,microtype,upgreek}
\usepackage[colorlinks=true,linkcolor=blue,citecolor=blue,urlcolor=blue]{hyperref}
\usepackage{url}
\usepackage{float}
\newcommand{\dd}{\mathrm{d}}

\begin{document}
\title{Stationary Action Principles for Neuronal Growth}
\author{Nathan Brodeur}
\affiliation{Department of Physics and Astronomy, Tufts University, Medford, MA 02155, USA}
\author{Cristian Staii}
\email{cstaii01@tufts.edu}
\affiliation{Department of Physics and Astronomy, Tufts University, Medford, MA 02155, USA}
\begin{abstract}
Neuronal growth is guided by biochemical, mechanical, and geometrical cues while remaining subject to stochastic fluctuations associated with cytoskeletal dynamics, adhesion, signaling, and growth cone sensing. Here, we develop
{two complementary} {variational descriptions of}
mechanically guided axonal growth. First, for neurons growing on periodic parallel micropatterns, we use previously established Langevin and Fokker--Planck descriptions to derive a stochastic path action that combines deterministic substrate guidance with random fluctuations. The probability of a growth trajectory decreases exponentially with this action, so that, for fixed boundary conditions, the most probable axonal trajectories correspond to minimum action paths. Second, for axons growing on substrates with spatially varying stiffness, we show that stiffness-dependent traction can generate {a mechanical guidance response}.  {At an interface between regions of different stiffness,} this {traction-dependent mechanical response produces axonal turning and provides} the basis for a {variational description of refraction}. In a quasi-static limit, the same mechanical framework leads to a Fermat-like spatial action with an effective axonal refractive index determined by the traction--stiffness relationship. This formulation predicts refraction of axons across stiffness boundaries and yields an approximate Snell-like relation and critical angle condition.  {The stochastic and Fermat-like actions have different physical meanings} and {ranges of validity, but both provide path-based descriptions of how substrate-}dependent
{mechanical cues constrain axonal trajectories.} {These results connect} measurable  {quantities
including drift, diffusion, substrate stiffness, traction, and growth direction} to {global properties of axonal paths} and suggest new strategies for designing biomimetic substrates for controlled neuronal growth and regeneration.
\end{abstract}
\maketitle
\section{Introduction}
\label{sec:intro}
The formation of neuronal networks depends on the ability of developing neurons to extend axons over long distances, select appropriate directions, and establish connections with target cells. Axons and dendrites form the principal neuronal processes, but axons are distinguished by their capacity for long-range growth and signal transmission~\cite{Azevedo2009, Striedter2016, Arimura2007}. Their extension is directed by the growth cone, a dynamic structure at the distal axon tip that integrates sensory information from the extracellular environment and converts it into cytoskeletal remodeling, adhesion, traction, and forward motion~\cite{Lowery2009,Huber2003,Franze2010,Franze2020, Medeiros2006, Athamneh2017}. Growth cones respond to diffusible and substrate-bound biochemical signals, cell adhesion molecules, electric fields, substrate mechanics, and topographical features~\cite{Goodhill2003,Franze2013,Moeendarbary2014, Kiryushko2004, Takano2019, Riveline2001}. These cues act on a cytoskeleton composed primarily of actin filaments and microtubules whose coordinated dynamics determine protrusion, turning, and axonal elongation~\cite{Samuels1996, Franze2009, Coles2015, Fivaz2007, Toriyama2010}. Cell--substrate coupling is mediated by adhesion complexes that connect the actin network to the surrounding material and regulate the conversion of retrograde actin flow into traction forces~\cite{Koch2012, Hyland2014, Polackwich2015,Jakobs2015, Kumarasinghe2022}. Axonal growth is therefore neither purely mechanical nor purely biochemical. It is an active, multiscale process in which sensing, force generation, transport, adhesion, and cytoskeletal dynamics continuously interact.

An additional difficulty is that these processes are intrinsically noisy. Stochastic receptor binding, molecular transport, actin and microtubule polymerization, adhesion turnover, signaling reactions, and filopodial activity generate fluctuations on time and length scales relevant to growth cone motion~\cite{DeGennes2007,DeRooij2017, pearson2011modeling, Oliveri2022, Betz2006, Simpson2009, Duswald2024, Staii2025}. Nevertheless, axons often display reproducible statistical responses to controlled guidance cues. This combination of variability at the level of individual trajectories and robust organization at the population level has motivated stochastic descriptions of neuronal growth based on Langevin and Fokker--Planck equations~\cite{Rizzo2013,Yurchenko2019,Vensi2019,Staii2023,Staii2024,Staii2025}. In these models, deterministic terms encode guidance and feedback, whereas stochastic terms describe fluctuations around the preferred motion. Such a separation is especially useful on engineered substrates, where mechanical and geometrical cues can be varied independently and the corresponding drift, diffusion, speed, orientation, and correlation parameters can be measured experimentally~\cite{Descoteaux2022,Staii2023,Teixeira2020}.

In our previous work, cortical neurons cultured on poly-D-lysine-coated {polydimethylsiloxane} (PDMS) substrates with periodic parallel micropatterns were found to preferentially extend axons along the substrate features~\cite{Basso2019,Yurchenko2021,Yurchenko2019,Vensi2019,Descoteaux2022,Staii2023}. The degree of alignment depends on the pattern spatial period and reflects a contact-guidance mechanism in which growth cone adhesion and traction couple neuronal motion to surface geometry~\cite{Descoteaux2022,Buskermolen2020}. The dynamics can be represented by nonlinear Langevin equations and associated Fokker--Planck equations with experimentally measurable deterministic alignment terms and stochastic diffusion coefficients~\cite{Yurchenko2019,Vensi2019,Staii2023,Staii2024}. A complementary mechanical description relates the preferred orientations to the bending cost of an axon following curved surface features~\cite{Descoteaux2022}. Similar stochastic guidance has also been observed on asymmetric textured and tilted nanostructured surfaces, where surface anisotropy produces directional bias in axonal growth~\cite{beighley2012alignment,Spedden2014Asymmetry,Basso2019}. These results establish a quantitative connection between substrate structure, mechanical interactions, deterministic guidance, and stochastic growth trajectories.

The existence of reproducible path selection in a noisy environment raises a broader theoretical question: can axonal trajectories be organized by a variational principle? In classical mechanics, a dynamical system with generalized coordinates $q_i$ can be described through a scalar Lagrangian $L(q_i,\dot q_i,t)$ and the action 
\begin{equation}
S[q]=\int_{t_i}^{t_f}L(q_i,\dot q_i,t)\,dt.
\label{eq:intro_action}
\end{equation}
The 
 physical trajectory makes the action stationary under variations with fixed endpoints,
\begin{equation}
\delta S=0,
\label{eq:intro_stationary}
\end{equation}
which gives the Euler--Lagrange equations
\begin{equation}
\frac{d}{dt}\left(\frac{\partial L}{\partial \dot q_i}\right)
-\frac{\partial L}{\partial q_i}=0.
\label{eq:intro_EL}
\end{equation}
The more precise terminology is therefore the principle of stationary action: depending on the problem, the stationary value may be a minimum, maximum, or saddle point. The Lagrangian formulation is valuable not only because it reproduces local differential equations, but because it replaces a set of local evolution laws by a global scalar constraint on admissible trajectories. When such a formulation exists, it can expose symmetries and conserved or energy-like quantities, provide alternative routes to stability analysis, and restrict the classes of dynamical laws that are compatible with a prescribed action. Conversely, determining whether a known set of differential equations admits a Lagrangian belongs to the inverse problem of the calculus of variations~\cite{Pham2023,NucciTamizhmani2012}.

Although the Lagrangian formalism originated in mechanics, variational approaches have increasingly been explored in biology. A recent review emphasizes that Lagrangians can be constructed for several population dynamics models, including Lotka--Volterra, Verhulst, Gompertz, host--parasite, and epidemic models, and that standard and non-standard Lagrangians can provide information about the dynamics without requiring direct solution of the original equations~\cite{Pham2023}. Earlier work developed canonical and Lagrangian descriptions of population kinetics and interacting species~\cite{Kerner1964,TrubatchFranco1974,NucciTamizhmani2012}. Stationary action concepts have also been applied to protein folding, where variations of an action were used to examine the stability of folding trajectories~\cite{SimmonsWeiner2013}, and to neuronal computation, where a least action principle has been proposed for real time learning in cortical circuits~\cite{Senn2024}. Related variational formulations include the free-energy principle for biological self-organization and inference~\cite{Friston2023}. These applications differ substantially in their physical interpretation, but together they illustrate a general strategy: complex biological dynamics can sometimes be recast in terms of a path functional whose stationarity or minimization constrains the observed evolution.

At the level of cell motility, the connection between variational ideas and stochastic dynamics is especially important. Migrating cells are not passive particles moving in equilibrium potentials. Their motion is generated by active cytoskeletal stresses, polarization, adhesion turnover, and force transmission to the substrate, and these processes are often represented through force-balance or stochastic continuum models~\cite{Alert2020}. For single cell motion, stochastic descriptions of chemotaxis and persistent migration have successfully separated systematic drift from fluctuating motion, as demonstrated, for example, in models of Dictyostelium chemotaxis and motility~\cite{amselem2012stochastic,li2011dicty}. This class of models provides a natural bridge to stochastic action formalisms because the probability of a full trajectory can be constructed from the same local transition probabilities that define the Langevin or Fokker--Planck dynamics. Recent minimum action proposals have also considered cellular development more directly, while related work has formulated adaptive Physarum dynamics in Lagrangian terms~\cite{Zandie2025,SolePlaMauri2025}. These studies are conceptually distinct from classical conservative mechanics, but they support the broader idea that path-based variational descriptions can be useful for active biological systems when the action is defined consistently with the underlying dynamics.

For cellular motion, variational and stochastic path descriptions are particularly relevant because cells are active, dissipative, and strongly coupled to fluctuating environments. Conventional models of cell migration commonly use force balance, polarity fields, active stresses, adhesion dynamics, or stochastic equations of motion rather than conservative Newtonian mechanics~\cite{Alert2020}. Recent work has nevertheless proposed minimum action descriptions of cellular development and adaptive biological motion~\cite{Zandie2025,SolePlaMauri2025}. {At the stochastic level, Wang and collaborators developed a path-information formulation in which an ensemble of possible trajectories is assigned probabilities according to an action}~\cite{Wang2005,Wang2009}. Maximizing path information subject to normalization and a constraint on the 
ensemble-averaged action gives
\begin{equation}
P_k=\frac{1}{Z}\exp\left(-\eta A_k\right),
\label{eq:intro_wang}
\end{equation}
where $P_k$ is the probability of the $k$th path, $A_k$ is the action associated 
with that path, $\eta$ is the Lagrange multiplier associated with the constraint 
on the mean action, and \mbox{$Z=\sum_k \exp(-\eta A_k)$} is the normalization factor. For positive $\eta$, paths with smaller action are 
therefore assigned larger probabilities~\cite{Wang2005,Wang2009}.

Exponential action dependence has also been derived directly for Langevin motion with Gaussian noise, including extensions intended for dissipative stochastic systems~\cite{ElKaabouchi2015}. Related work connects generalized path information to nonlinear Fokker--Planck equations and anomalous diffusion~\cite{LiWang2008}. More generally, the Onsager--Machlup and Freidlin--Wentzell formalisms assign path probabilities to noisy dynamical systems through functionals that penalize deviations from deterministic drift~\cite{OnsagerMachlup1953,FreidlinWentzell2012}. These frameworks are particularly appropriate for biological motility because the relevant action is a statistical path cost rather than the conservative mechanical action $\int (T-V)dt$.

The present work applies these variational ideas to two complementary forms of
mechanically guided axonal growth. For neurons growing on periodic parallel
micropatterns, we begin from experimentally calibrated Langevin and
Fokker--Planck models~\cite{Rizzo2013,Yurchenko2019,Vensi2019,Descoteaux2022,
Staii2023,Staii2024,Staii2025} and derive a stochastic path action from the
Gaussian transition probabilities of the growth cone dynamics. The resulting
trajectory probability decreases exponentially with action, consistent with
the broader Onsager--Machlup, Freidlin--Wentzell, and path-information
frameworks discussed above~\cite{OnsagerMachlup1953,FreidlinWentzell2012,
Wang2005,Wang2009}. For fixed boundary conditions, the most probable axonal
trajectories are therefore minimum action paths, where the action measures the
cumulative deviation from the experimentally determined deterministic guidance
field rather than a conservative mechanical energy.

A complementary variational description is obtained for axons crossing a
mechanical interface between PDMS regions of different stiffness. Building on
molecular clutch descriptions of stiffness-dependent traction
\cite{ChanOdde2008,Sens2013,Bangasser2013} and mechanical models of axonal
durotaxis~\cite{Oliveri2021,Kassianides2025}, we show that the quasi-static
growth cone and axonal bending response defines an effective axonal refractive
index. A Fermat-like stationary action then leads to a Snell-like relation for
axonal refraction at the stiffness boundary. The experimentally observed
changes in trajectory direction across the two region PDMS interface provide a
direct test of this description and connect measurable quantities such as
substrate stiffness, growth cone traction, axonal bending, and refraction
angles. Although the stochastic and Fermat-like actions arise from different
levels of description, both express how substrate-dependent mechanical
interactions constrain the paths available to a growing axon.

These results are also relevant to the broader development of engineered
environments for neuronal guidance. Advances in microfabrication,
microfluidics, and mechanically patterned culture substrates increasingly
allow geometric, biochemical, and mechanical guidance cues to be controlled
independently~\cite{Teixeira2020,kundu2013superimposed,gladkov2017design,
Hyland2014,Jurchenko2015,ishihara2011assay,Descoteaux2022,
Spedden2012Elasticity,Koch2012,CastroDominguez2023,Pfister2011}. Such
platforms provide experimental systems in which theoretical predictions for
alignment, turning, refraction, drift, and stochastic variability can be
tested quantitatively. More broadly, understanding how substrate geometry and
mechanics regulate axonal path selection is important for the design of
growth-permissive scaffolds and biomimetic interfaces intended to promote
directed axonal extension, nerve repair, and functional reconnection
\cite{Pfister2011,Holland2015,Bayly2014,DeRooij2018a,DeRooij2018b,
Mahar2018,Teixeira2020,Ahmadzadeh2015,Montanino2018}.

The principal significance of the present work is therefore to place two
experimentally accessible forms of axonal guidance within a common
path-based framework. Substrate geometry and stiffness determine measurable
mechanical guidance responses, while stochastic cellular processes generate
ensembles of possible trajectories around those responses. The stochastic
action developed for parallel micropatterns quantifies the probability of
these trajectories, whereas the Fermat-like action provides a reduced spatial
description of stiffness-dependent refraction. Together, these results extend
{variational action principles} from general stochastic and biological systems to
quantitative models of axonal guidance and provide experimentally testable
relations linking substrate design, cellular mechanics, and neuronal growth
trajectories.

\section{Materials and Methods 
}
\label{sec:methods}

\textit{Neuronal Cell Culture}. 
Cortical neurons were prepared from embryonic day 18 rat embryos. Procedures involving animal tissue were approved by the Tufts University Institutional Animal Care and Use Committee and performed in compliance with NIH guidelines for the Care and Use of Laboratory Animals. Cell dissociation and culture were carried out according to protocols established in our previous work~\cite{Rizzo2013,Spedden2012Elasticity,Basso2019,Kumarasinghe2022,Staii2023,Staii2024,Yurchenko2021,Yurchenko2019,Vensi2019,Descoteaux2022}. {Briefly, the isolated cortices were incubated in }$5$~mL{ of trypsin at }
$37^\circ \mathrm{C}${ for }$20$~min{, after which enzymatic digestion was
quenched with }$10$~mL{ of soybean trypsin inhibitor. The tissue was
mechanically dissociated and centrifuged, and the resulting cell pellet was
resuspended in }$20$~mL{ of Neurobasal medium supplemented with GlutaMAX,
B27, and penicillin--streptomycin}
\cite{Rizzo2013,Spedden2012Elasticity,Basso2019,Kumarasinghe2022,
Yurchenko2021,Yurchenko2019,Vensi2019,Descoteaux2022}. Previous immunostaining measurements confirmed the high neuronal purity of these cultures~\cite{Spedden2012Elasticity}. Cells were plated on poly-D-lysine (PDL)-coated polydimethylsiloxane (PDMS) substrates at a density of approximately 4000 cells/cm$^2$. Before cell plating, the substrates were coated with PDL (0.1 mg/mL; Sigma-Aldrich, St. Louis, MO, USA). Low-density neuronal cultures in the range of approximately 3000--7000 cells/cm$^2$ promote the formation of long, spatially separated axons and are therefore well suited for quantitative measurements of growth cone trajectories and axonal guidance~\cite{Basso2019,Kumarasinghe2022,Staii2023,Staii2024,Yurchenko2021,Yurchenko2019,Vensi2019,Descoteaux2022}.

\begin{figure*}[t]
\centering
    \includegraphics[width=0.85\textwidth]{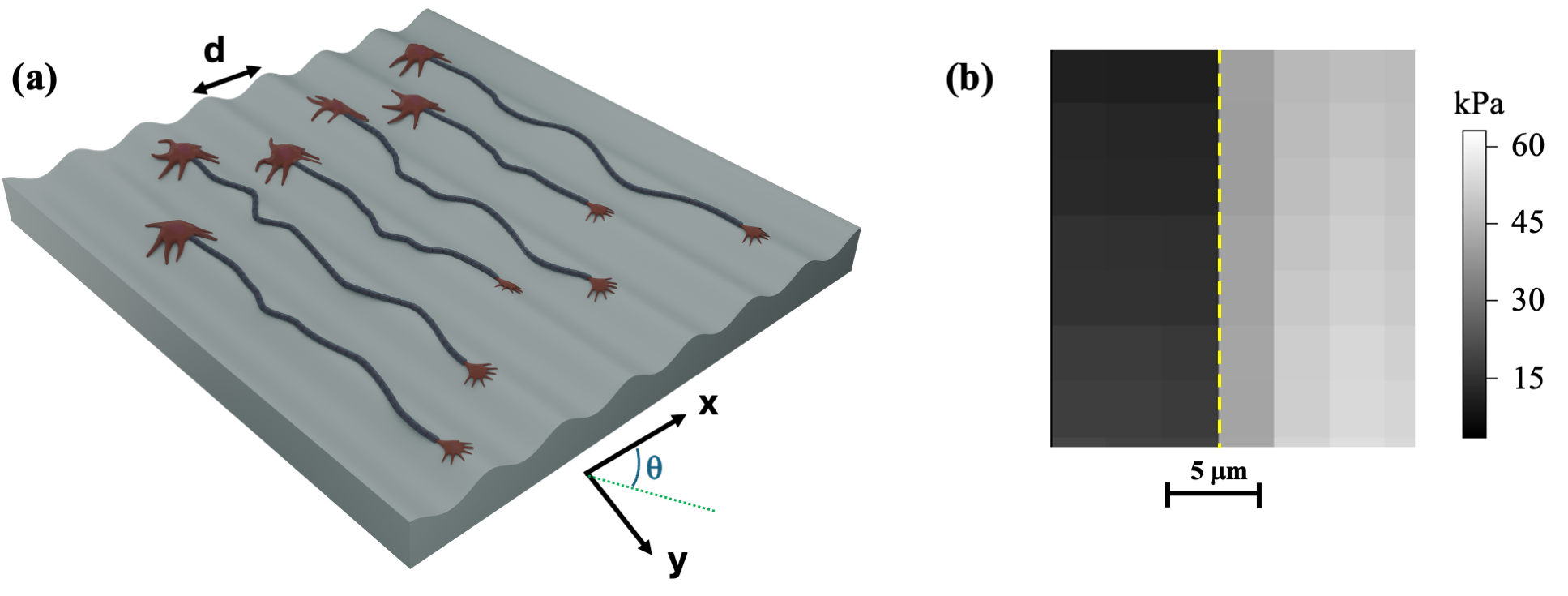}
    \caption{(\textbf{a}) Schematic 
 representation of neuronal growth on a PDMS substrate containing periodic parallel micropatterns. The surface patterns are periodic in the $x$ direction with spatial period $d$, while the $y$-axis is defined parallel to the ridges and grooves. A cortical neuron is shown extending an axon preferentially along the pattern direction. The instantaneous growth angle $\theta(t)$ is defined as the angle between the growth cone {displacement vector measured between two consecutive frames} and the $x$-axis. (\textbf{b}) Representative AFM force maps obtained from the two mechanically distinct regions of a stiffness-patterned PDMS substrate. Analysis of the AFM force measurements yielded Young's moduli of $8 \pm 2$~kPa for the softer region and $57 \pm 6$~kPa for the stiffer region. The boundary between the two regions is shown schematically by the dotted yellow line.}
    \label{fig1}
\end{figure*}

\textit{PDMS Substrates with Parallel Micropatterns}. The geometrically patterned substrates consisted of parallel PDMS ridges separated by grooves, as illustrated schematically in Figure~\ref{fig1}a. The patterns were fabricated using the imprinting procedure employed in our previous studies~\cite{Basso2019,Yurchenko2021,Yurchenko2019,Vensi2019,Descoteaux2022,Staii2023,Staii2024}. {The PDMS solution was poured over diffraction gratings with slit
separations ranging from }$1$--$10~\upmu$m{ in increments of }$1~\upmu$m
{ and a total surface area of }$25\times25~\mathrm{mm}^2$
{(Newport Corp., Irvine, CA, USA). The PDMS films were
allowed to polymerize for }$48$~h{ at room temperature, then peeled from
the diffraction gratings and cured at } $55^\circ \mathrm{C}$ { for }$3$~h
\cite{Yurchenko2021,Yurchenko2019,Vensi2019,Descoteaux2022}. The $y$-axis was defined along the direction of the ridges and grooves, while the $x$-axis was defined perpendicular to the pattern direction. {Before neuronal plating, the patterned surfaces were spin-coated with }
$3$~mL{ of }$0.1$~mg/mL{ PDL for }$10$~min{ at }$1000$~rpm{
and sterilized by ultraviolet exposure for }$30$~min
\cite{Yurchenko2021}.

The characteristic dimensions of the parallel micropatterns were verified by atomic force microscopy (AFM). These substrates provide a controlled geometrical cue that promotes preferential alignment of axons along the ridge direction through contact guidance. Previous measurements demonstrated that the degree of alignment depends on the spatial period $d$ and is strongest when the characteristic dimensions of the surface pattern are comparable to those of the growth cone~\cite{Basso2019,Yurchenko2021,Yurchenko2019,Vensi2019,Descoteaux2022,Staii2023}.

\textit{PDMS Substrates with Regions of Different Stiffness}. A second class of PDMS substrates was fabricated to provide a spatially varying mechanical cue while minimizing surface topographical variations. PDMS stiffness was varied by changing the ratio of elastomer base to curing agent, which modifies the crosslink density of the polymer network. Previous studies have demonstrated the fabrication of PDMS substrates containing adjacent regions with different elastic moduli and their use for investigating cellular responses to spatial variations in substrate mechanics~\cite{Gray2003,Chou2009}. In particular, composite PDMS substrates can be prepared by sequentially curing formulations with different elastic properties to generate neighboring regions of distinct stiffness~\cite{Chou2009}. For the substrates used here, two PDMS formulations with different base to curing agent ratios were prepared separately to obtain a softer and a stiffer region. Each formulation was thoroughly mixed and degassed before casting. {Following the general sequential fabrication procedure described in Ref.}~\cite{Chou2009}, {the
first PDMS formulation was introduced into one region of the substrate and
partially cured at }$65^\circ\mathrm{C}$
{ for }$24${ to stabilize the region and
establish a well-defined boundary. The second PDMS formulation was then
introduced adjacent to the first region, and the complete two-region substrate
was cured at room temperature for} $48$~h{. This sequential
curing procedure produced a mechanically continuous PDMS substrate containing
neighboring regions of distinct elastic modulus separated by an approximately
straight boundary,} shown schematically by the dotted yellow line in Figure~\ref{fig1}b.

The mechanical properties of the two regions were characterized by AFM force mapping. Representative maps are shown in Figure~\ref{fig1}b. Analysis of the AFM force measurements yielded a Young's modulus of $E_{s,1}=57\pm6$~kPa
for the stiffer region and $E_{s,2}=8\pm2$~kPa for the softer region, corresponding to an approximately sevenfold difference in substrate stiffness. After mechanical characterization, the complete substrate was subjected to the same surface treatment and uniformly coated with PDL before neuronal plating, thereby maintaining comparable cell-contacting surface conditions across the two regions. The resulting two-region geometry provides a well-defined mechanical interface for investigating changes in axonal growth direction produced by a discrete variation in substrate stiffness. Individual axons can approach the interface from either region and subsequently cross into the second region, change direction near the boundary, or remain predominantly within the initial region. These trajectories provide the experimental basis for comparing stiffness-dependent changes in axonal growth with the {Fermat-like stationary action} description developed in this work.

\textit{Imaging and Data Acquisition}. AFM and fluorescence microscopy were used to characterize the substrates and neuronal growth. AFM measurements were performed using an MFP-3D system (Asylum Research) equipped with a BioHeater fluid cell and integrated with an inverted Nikon Eclipse Ti microscope (Micro Video Instruments, Avon, MA, USA). {AFM force measurements were analyzed using the calibration and fitting
procedures established in our previous work}~\cite{Spedden2012Elasticity}{.
Measurements were performed using Olympus Biolever cantilevers (Asylum
Research) with a nominal spring constant of }$0.03~\mathrm{N/m}${. Before
each set of measurements, the cantilever was calibrated in air on clean glass
to determine the optical lever sensitivity and spring constant, and recalibrated
in the measurement medium using both thermal calibration and
deflection-versus-}$z${ curves acquired on a clean glass region. Force curves
were acquired at a cantilever }$z${ velocity of approximately }
$2~\upmu\mathrm{m\,s^{-1}}${, and Young's modulus values were obtained by
fitting the force--indentation curves with the Hertz--Sneddon contact model for
a conical AFM tip.} 

\begin{figure*}[t]
\centering
    \includegraphics[width=0.85\textwidth]{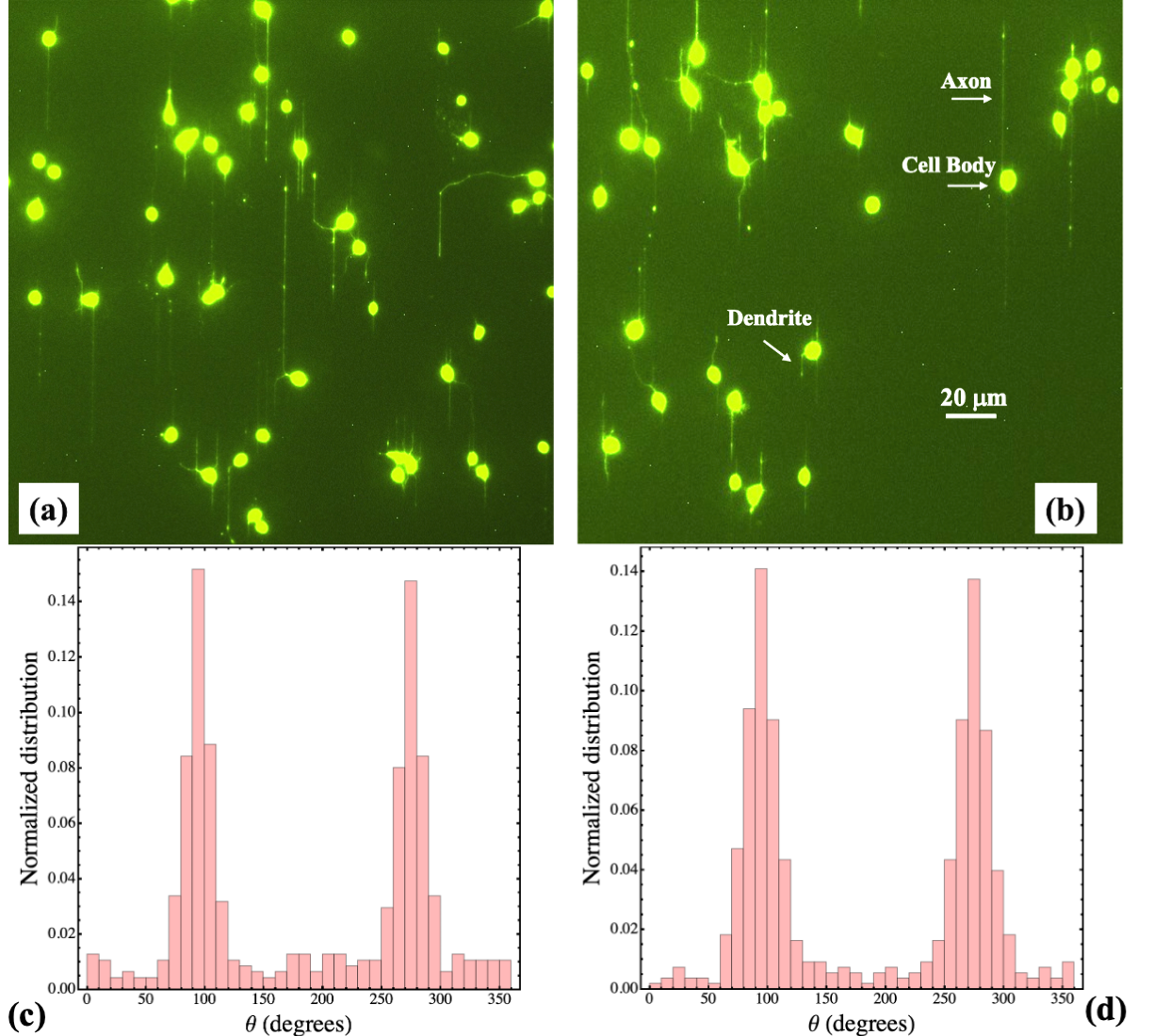}
  \caption{Representative 
 fluorescence images of cortical neurons growing on PDMS substrates containing parallel micropatterns with spatial periods $d=4~\upmu$m in (\textbf{a}) and $d=6~\upmu$m in (\textbf{b}). Both images were obtained at $t=50$~h after plating. Axonal growth occurs preferentially along the direction of the parallel ridges and grooves, illustrating the directional guidance produced by the substrate geometry. Major structural components of the neurons, including the cell body, axons, and dendrites, are also indicated. The scale bar shown in (\textbf{b}) applies to both images. (\textbf{c},\textbf{d}) Representative normalized distributions of axonal growth angles for neurons growing on the $d=4~\upmu$m and $d=6~\upmu$m micropatterns, respectively. {The distributions were constructed from $N=786$ different axon segments analyzed for $d=4~\upmu$m, and $N=821$ segments for $d=6~\upmu$m, respectively}. Both distributions exhibit pronounced peaks near $\theta=90^\circ$ and $270^\circ$, corresponding to the preferred growth directions parallel to the micropatterns under the angular convention defined in Figure~\ref{fig1}a. The strong angular localization illustrates the directional alignment of axonal growth induced by the substrate geometry.}  
     \label{fig2}
\end{figure*}

For the parallel micropatterned substrates, AFM topographic images were used to characterize the ridge height, spatial period, and surface morphology. For the substrates containing regions of different stiffness, AFM topographic imaging was used to characterize the surface near the interface, while AFM force mapping was used to determine the local elastic modulus within each region. {Fluorescence imaging was used to identify neuronal processes and monitor axonal growth (Figure}~\ref{fig2}a,b). {For imaging, live cultures were rinsed with PBS and incubated
for }$30$~min{ at } $37^\circ \mathrm{C}$ { in PBS containing }$50$~nM
{ Tubulin Tracker Green (Oregon Green 488 Taxol, bis-Acetate; Life
Technologies). Samples were then rinsed twice with PBS and transferred to
fresh PBS for imaging}
\cite{Yurchenko2021,Descoteaux2022}. Images were acquired using the inverted microscope with standard FITC filters, with excitation and emission wavelengths of approximately 495 nm and 521 nm, respectively. Individual growth cones were imaged at successive time points to determine their position, velocity, and direction of motion. Particular attention was given to axons that were sufficiently isolated from neighboring cells and neuronal processes to permit unambiguous identification and tracking of the corresponding growth cone throughout the observation period.

\textit{Data Analysis}. Growth cone trajectories were analyzed using ImageJ (version 1.54k, 
NIH). {The position of each growth cone was determined from the center of the
growth cone in successive fluorescence images, following the tracking procedure
used in our previous studies}
\cite{Yurchenko2021,Descoteaux2022}. The two-dimensional position of each growth cone was represented by $\mathbf{r}(t)=[x(t),y(t)]$ and determined from sequential fluorescence images. Images were acquired at intervals of $\Delta t=5$~min over a total observation period of 1 hr. Measurements were performed at multiple times after plating, including $t=10$, 15, 20, 25, 30, 35, 40, 45, and 50 h. The 5 min sampling interval was chosen so that the growth cone displacement between successive images exceeded the experimental spatial resolution of approximately $0.1~\upmu$m while remaining sufficiently short for $\Delta\mathbf{r}/\Delta t$ to provide a local estimate of the instantaneous growth velocity. For successive growth cone positions $\mathbf{r}(t_i)$ and $\mathbf{r}(t_{i+1})$, the displacement was calculated as $\Delta\mathbf{r}_i=\mathbf{r}(t_{i+1})-\mathbf{r}(t_i)$ and the corresponding velocity as $\mathbf{V}_i=\Delta\mathbf{r}_i/\Delta t$. For neurons growing on the parallel micropatterns, the instantaneous growth angle $\theta$ was defined relative to the $x$-axis (Figure~\ref{fig1}a), such that $\theta_i=\tan^{-1}(\Delta y_i/\Delta x_i)$, and mapped onto the interval $0\leq\theta<2\pi$. With this convention,
$\theta=90^\circ$ and $270^\circ$ correspond to growth parallel to the
micropattern direction (Figure~\ref{fig2}c,d). {The interval} $0\leq\theta<2\pi$ {was divided into 36 bins of width}
$\Delta\theta=10^\circ$. The resulting trajectories were used to determine distributions of growth angles and velocities and to calculate the statistical quantities entering the Langevin and Fokker--Planck descriptions, including effective drift and diffusion parameters. 

To determine the angular distributions shown in
Figure~\ref{fig2}c,d, individual axons were tracked using ImageJ (version 1.54k) 
 and divided
into consecutive segments of approximately $20~\upmu$m in length, following
the procedure used in our previous studies~\cite{Vensi2019, Descoteaux2022}. {Experimental measurements showed that, over length scales of approximately
$20~\upmu$m, individual axonal trajectories can be approximated accurately by
straight line segments. This observation motivated the choice of segment
length used for the angular analysis.} The orientation of each segment
was measured relative to the $x$-axis using the same angular convention
defined above. The full angular range $0\leq\theta<2\pi$ was divided into
equal width intervals, and the number of axonal segments within each interval
was used to construct the corresponding angular distributions. The
distributions were normalized by the total number of analyzed segments.

For substrates containing two regions of different stiffness, the normal to the interface was used as the reference direction for analyzing changes in axonal trajectory. The incident angle $\alpha_1$ was determined from the direction of growth immediately before the growth cone reached the stiffness boundary, while the transmitted angle $\alpha_2$ was determined from the trajectory after the axon entered the second region. Both angles were measured relative to the interface normal. Individual trajectories were classified according to whether the axon crossed the boundary, changed direction near the interface, or continued to grow predominantly within the initial stiffness region. For axons crossing the boundary, the measured pairs $(\alpha_1,\alpha_2)$ were used to characterize stiffness-dependent changes in growth direction and to compare the experimental trajectories with the predicted refraction relation. 

\section{Results}
\label{sec:results}

\subsection{Axonal Growth on Periodic Micropatterns}
\label{sec:axon_model}

Figure~\ref{fig2} shows representative fluorescence images of cortical neurons growing on a PDMS substrate containing parallel micropatterns with spatial period $d=4~\upmu$m (Figure~\ref{fig2}a) and $d=6~\upmu$m (Figure~\ref{fig2}b). Both images were obtained at $t=50$~h after plating. Axons preferentially extend along the direction of the parallel micropatterns, demonstrating the strong contact guidance produced by the substrate geometry. This behavior is consistent with our previous measurements of neuronal growth on periodically patterned PDMS substrates~\cite{Yurchenko2019,Vensi2019,Yurchenko2021,Descoteaux2022,Staii2023,Staii2025}.

The stochastic dynamics underlying this alignment can be described using the spatial Fokker--Planck model developed for neuronal growth on parallel micropatterns~\cite{Descoteaux2022,Staii2025}. The probability density $p(\mathbf r,t)$ for the growth cone position satisfies
\begin{equation}
\frac{\partial p}{\partial t}
=
D\nabla^2p
+
\frac{1}{\gamma}
\nabla\cdot
\left[
p\nabla U_{\mathrm{eff}}(\mathbf r,t)
\right],
\label{eq:spatial_fp}
\end{equation}
where $D$ is an effective motility coefficient, $\gamma$ is a damping coefficient, and $U_{\mathrm{eff}}$ represents the effective guidance potential~\cite{Descoteaux2022}. Axonal growth is an active nonequilibrium process, so $D$ is an effective stochastic cell motility parameter. Experiments on micropatterned PDMS substrates yield representative values $D\simeq21~\upmu$m$^2$ h$^{-1}$ and $\gamma=(0.11\pm0.04)$ h$^{-1}$~\cite{Descoteaux2022, Staii2023}.

For the low neuronal densities considered here, cell to cell interactions can be neglected to first approximation, and the effective potential can be written as the sum of substrate and feedback contributions, $U_{\mathrm{eff}}=U_{\mathrm{sub}}+U_{\mathrm{fb}}$. Previous analysis of growth on parallel PDMS micropatterns showed that the transverse substrate contribution can be represented by a periodic potential and the feedback response by a moving harmonic contribution~\cite{Descoteaux2022,Staii2025},
\begin{equation}
\begin{aligned}
U_{\mathrm{sub}}(x)&=U_0\sin\!\left(\frac{2\pi x}{d}\right),\\
U_{\mathrm{fb}}(x,t)&=U_F[x-\langle x(t)\rangle]^2,
\end{aligned}
\label{eq:micro_potentials}
\end{equation}
where $x$ is the coordinate perpendicular to the parallel ridges (Figure~\ref{fig1}a), $d$ is the pattern spatial period, and $U_0$ and $U_F$ characterize the strengths of the substrate and feedback contributions, respectively. This effective potential is a coarse-grained description of the observed guidance and should not be identified directly with a conservative mechanical energy.

The corresponding Fokker--Planck equation is equivalent to the overdamped Langevin equation~\cite{Descoteaux2022,Staii2025, li2011dicty}
\begin{equation}
\begin{aligned}
dx&=b(x,t)\,dt+\sqrt{2D}\,dW_t,\\
b(x,t)&=-\frac{1}{\gamma}\frac{\partial U_{\mathrm{eff}}}{\partial x},
\end{aligned}
\label{eq:spatial_sde}
\end{equation}
where $W_t$ is a standard Wiener process representing Gaussian stochastic fluctuations, with increments of zero mean and variance $dt$. The growth cone trajectory therefore results from a deterministic drift imposed by substrate guidance and cellular feedback together with stochastic fluctuations associated with the active growth process.

{The reduction to the transverse coordinate }$x${ is motivated by the geometry
of the parallel micropatterns }(Figure~\ref{fig1}a){. To leading order, the substrate
guidance potential varies periodically across the ridges but is approximately
translationally invariant along the ridge direction }$y${. The resulting
one-dimensional Fokker--Planck description therefore characterizes the
transverse localization and crossing dynamics of the growth cones rather than
their full two-dimensional trajectories. Longitudinal advance and correlations between speed and orientation are not represented explicitly in this reduced description.}

The directional response of axons can also be described by stochastic dynamics in orientation space. The experimentally measured angular distributions are strongly peaked around directions parallel to the substrate patterns (Figure~\ref{fig2}c,d). {The angular measurements are used here to characterize and classify trajectory
orientation} {The angle} $\theta$ {is not included as an independent stochastic state variable in the action formulated only in terms of position.} The corresponding angular Fokker--Planck description contains a deterministic alignment drift characterized by an angular relaxation rate \(\gamma_\theta\), together with an angular diffusion coefficient \(D_\theta\)~\cite{Staii2025}. Near each preferred alignment direction, the angular stochastic dynamics may be written generically as
$$ \dd\theta = b_\theta(\theta)\dd t + \sqrt{2D_\theta}\,\dd W^\theta_t, $$
where \(b_\theta(\theta)\) is the substrate-induced orientational drift and \(W_t^\theta\) is a Wiener process representing angular fluctuations. Near a stable preferred growth direction \(\theta_\ast\), the drift can be linearized as
 $b_\theta(\theta)\simeq -\gamma_\theta(\theta-\theta_\ast)$,
showing explicitly that the deterministic dynamics tend to restore the growth direction toward alignment with the substrate pattern. Experimentally, the drift vanishes along the preferred directions and acts to rotate misaligned growth cones toward these stable orientations~\cite{Vensi2019,Staii2023,Staii2025}.

\subsubsection{Path Probability and Stochastic Action}

The Langevin description provides a direct connection between the experimentally measured stochastic dynamics and a minimum action formulation. For a trajectory sampled at intervals $\Delta t$, let $x_n=x(t_n)$ and $\Delta x_n=x_{n+1}-x_n$. Using an
Euler--Maruyama discretization of Equation~\eqref{eq:spatial_sde},
the short-time conditional transition probability is
\cite{OnsagerMachlup1953,FreidlinWentzell2012}
\begin{equation}
\begin{aligned}
P(x_{n+1}\mid x_n)
&=\frac{1}{\sqrt{4\pi D\Delta t}}
\\
&\quad\times\exp\!\left[
-\frac{[\Delta x_n-b(x_n,t_n)\Delta t]^2}
{4D\Delta t}\right].
\end{aligned}
\label{eq:transition_probability}
\end{equation}
Because the dynamics is Markovian, the probability of a discrete trajectory \mbox{$\Gamma=\{x_0,x_1,\ldots,x_N\}$} is the product of the transition probabilities. Apart from normalization factors, it can therefore be expressed as
\begin{equation}
\mathcal P[\Gamma]
\propto
\exp[-S_N(\Gamma)],
\label{eq:path_probability}
\end{equation}
where the discrete stochastic action is
\begin{equation}
S_N(\Gamma)
=
\sum_{n=0}^{N-1}
\frac{
\left[
\Delta x_n-b(x_n,t_n)\Delta t
\right]^2
}{
4D\Delta t
}.
\label{eq:discrete_action}
\end{equation}
This expression is particularly useful for experimental trajectory analysis because it can be evaluated directly from time-lapse growth cone positions once the drift $b(x,t)$ and motility coefficient $D$ have been determined. For two trajectories measured under the same conditions and with the same temporal discretization, Equation~\eqref{eq:path_probability} gives~\cite{OnsagerMachlup1953,FreidlinWentzell2012}
\begin{equation}
\frac{\mathcal P[\Gamma_1]}
{\mathcal P[\Gamma_2]}
=
\exp\left[
-\left(S_N[\Gamma_1]-S_N[\Gamma_2]\right)
\right].
\label{eq:path_ratio}
\end{equation}
Consequently, within this discretized stochastic description, a
trajectory with smaller stochastic action has a larger likelihood.
For specified initial and final conditions, the most probable
trajectory is therefore identified with the trajectory that minimizes
the discrete stochastic action.

In the continuum description, the corresponding Freidlin--Wentzell rate functional~\cite{FreidlinWentzell2012} associated with
Equation~\eqref{eq:spatial_sde} can be written as (derivation provided in Appendix~\ref{app:fw_derivation}):
\begin{equation}
S_{\mathrm{micro}}[x]
=
\frac{1}{4D}
\int_{t_i}^{t_f}
\left[
\dot{x}
+
\frac{1}{\gamma}
\frac{\partial U_{\mathrm{eff}}}{\partial x}
\right]^2dt,
\label{eq:micro_action}
\end{equation}
where we have used
$b(x,t)=-(1/\gamma)\partial U_{\mathrm{eff}}/\partial x$.

{The discrete likelihood in}
Equation~\eqref{eq:discrete_action}
{ follows directly from the assumed finite-dimensional Gaussian--Markov
transition model and is applicable without invoking an asymptotically
small-noise limit. The continuum expression in}
Equation~\eqref{eq:micro_action}
{ provides the corresponding leading Freidlin--Wentzell rate functional
and is interpreted in its standard asymptotic sense}~\cite{FreidlinWentzell2012}. {For the
experimental trajectories considered here, which are measured at finite
temporal and spatial resolution, the discrete likelihood provides the primary
description used for quantitative analysis, while the continuum functional
provides the corresponding variational description of the stochastic
dynamics.}

{More generally, for an isotropic two-dimensional diffusion process }
$d\mathbf r=\mathbf b(\mathbf r,t)\,dt+\sqrt{2D}\,d\mathbf W_t${, the
corresponding Freidlin--Wentzell rate functional has the form }
$S_{\mathrm{2D}}[\mathbf r]=(1/4D)\int_{t_i}^{t_f}
|\dot{\mathbf r}-\mathbf b(\mathbf r,t)|^2dt${. Equation~}\eqref{eq:micro_action}
{ is the one-dimensional transverse reduction of this more general
description, appropriate when the dominant substrate-induced guidance
variation occurs across the parallel micropatterns in the }$x${ direction} (Figure~\ref{fig1}a).

\begin{figure}[t]
    \includegraphics[width=0.95\columnwidth]{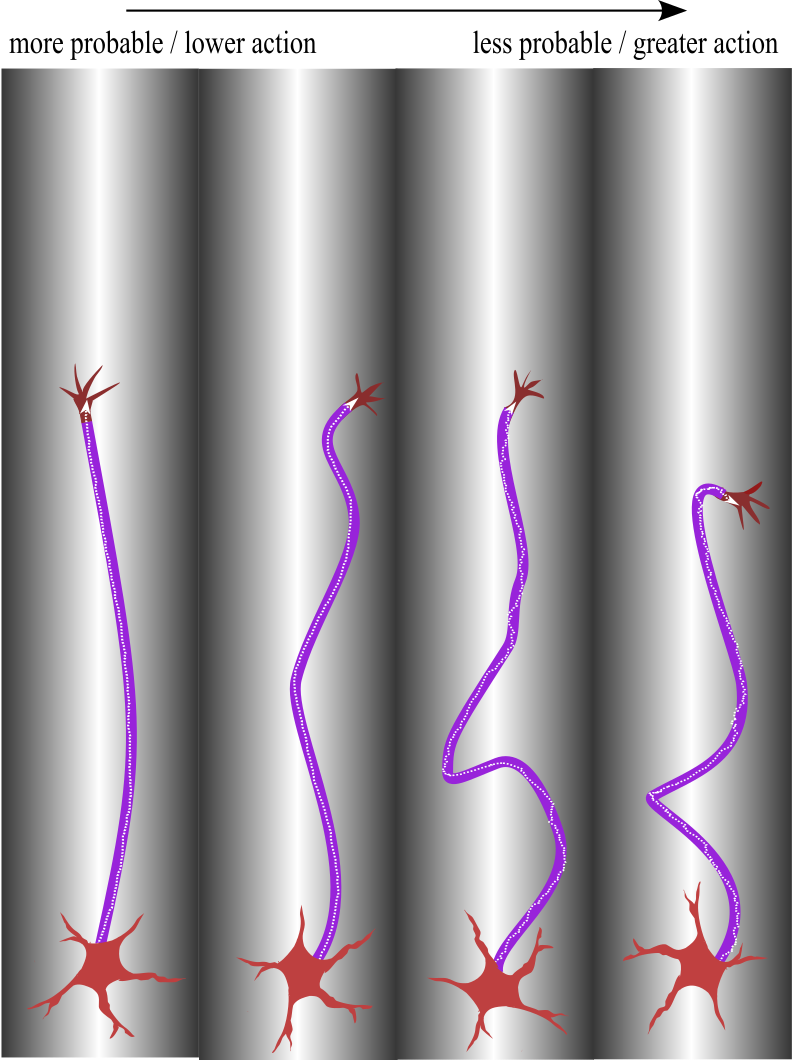}
    \caption{Schematic 
 illustration of the stochastic action picture for axonal growth on a periodic micropatterned substrate. Several possible noisy growth trajectories  are shown as dotted lines for axons originating from similar initial regions. Trajectories that remain approximately aligned with the substrate-preferred direction and closely follow the deterministic guidance drift have smaller values of the stochastic action and are therefore more probable. Increasing deviations from the deterministic drift, including stronger transverse excursions across the micropatterns, require larger stochastic fluctuations and correspond to progressively larger action and lower path probability. The figure illustrates the statistical interpretation of the minimum action principle: lower action paths are exponentially favored within the ensemble of possible axonal trajectories.}
     \label{fig3}
\end{figure}

The physical interpretation of Equation~\eqref{eq:micro_action} is statistical: among paths satisfying
the same boundary conditions, paths with smaller values of
$S_{\mathrm{micro}}$ are exponentially favored in the corresponding
rate-functional description. This does not mean that an individual
axon minimizes a conservative mechanical action. Rather, trajectories
that remain close to the deterministic guidance field require less
stochastic forcing and therefore have lower path cost and higher
probability. A trajectory that closely follows the deterministic drift
has a small value of $\dot{x}-b(x,t)$ and therefore a small stochastic
action. A trajectory that repeatedly moves against the deterministic
guidance field requires larger stochastic fluctuations and accumulates
a larger action. Thus, the minimum action trajectory corresponds to
the path requiring the smallest cumulative stochastic deviation from
the deterministic growth law. This formulation is natural for our
micropattern experiments because the measured dynamics is already
expressed as deterministic drift plus noise. This stochastic action picture is illustrated schematically in Figure~\ref{fig3}, where trajectories that remain aligned with the substrate-preferred direction have lower action and higher probability than paths exhibiting progressively larger transverse deviations.

Equation~\eqref{eq:micro_action} represents the leading
Freidlin--Wentzell rate functional. An exact Onsager--Machlup
functional for a nonlinear drift can contain an additional
drift-divergence term whose precise form depends on the definition of
the continuum path measure~\cite{OnsagerMachlup1953}. For direct
comparison with experimentally sampled growth cone trajectories,
Equation~\eqref{eq:discrete_action} therefore provides a particularly
direct likelihood-based measure of path cost.

\subsubsection{Minimum Action for the Micropatterned Growth Model}

Substitution of the experimentally motivated substrate and feedback potentials in Equation~\eqref{eq:micro_potentials} into the stochastic action gives~\cite{Descoteaux2022,Staii2025}

\begin{equation}
\begin{aligned}
S_{\mathrm{micro}}[x]
&=\frac{1}{4D}\int_{t_i}^{t_f}
\Bigg[\dot{x}+\frac{1}{\gamma}\Bigg(
\frac{2\pi U_0}{d}\cos\!\left(\frac{2\pi x}{d}\right)
\\
&\qquad\quad+2U_F[x-\langle x(t)\rangle]
\Bigg)\Bigg]^2\,dt.
\end{aligned}
\label{eq:explicit_spatial_action}
\end{equation}
Importantly, Equation~\eqref{eq:explicit_spatial_action} introduces no additional phenomenological guidance force. Its deterministic terms are the same substrate and feedback contributions already inferred from the Fokker--Planck description of the experimental trajectories~\cite{Descoteaux2022,Staii2025}. This equation has a direct interpretation. Axonal trajectories that remain localized near preferred regions of the parallel pattern and follow the deterministic alignment drift accumulate relatively small action. In contrast, persistent transverse motion across the micropatterns or motion opposing the deterministic drift requires larger stochastic fluctuations and produces a larger action. The experimentally observed concentration of axonal growth along the parallel patterns can therefore be interpreted as concentration of the trajectory ensemble around lower action paths. This interpretation is consistent with the dependence of axonal alignment on micropattern geometry. Previous experiments showed that directional guidance is strongest when the spatial period of the substrate is comparable to the characteristic size of the growth cone~\cite{Yurchenko2019,Vensi2019,Yurchenko2021,Descoteaux2022}. In the present framework, changes in pattern spacing modify the deterministic guidance landscape and therefore change the relative stochastic action of aligned and transverse trajectory families.

The stochastic action formalism can be tested directly using experimentally measured growth cone trajectories. For each micropattern spatial period \(d\), trajectories are sampled at the same time interval \(\Delta t\), and the discrete stochastic action \(S_N\) is calculated from Equation~\eqref{eq:discrete_action} using the measured increments \(\Delta x_n\), together with the deterministic drift \(b(x_n,t_n)=-(1/\gamma)\partial U_{\mathrm{eff}}/\partial x\) and the corresponding parameters obtained from the Fokker--Planck analysis for that substrate~\cite{Descoteaux2022,Staii2025}. The degree of directional alignment of an individual trajectory is quantified by \(A=-N^{-1}\sum_{n=1}^{N}\cos(2\theta_n)\). The factor \(2\theta\) is used because the parallel micropatterns possess directional symmetry: growth along a ridge in the directions \(\theta\) and \(\theta+\pi\) represents the same degree of alignment with the substrate, whereas \(\cos\theta\) would assign opposite signs to these two physically equivalent directions. Values of \(A\) approaching unity therefore identify trajectories that remain predominantly parallel to the substrate pattern.

\begin{figure*}[t]
\centering
    \includegraphics[width=0.85\textwidth]{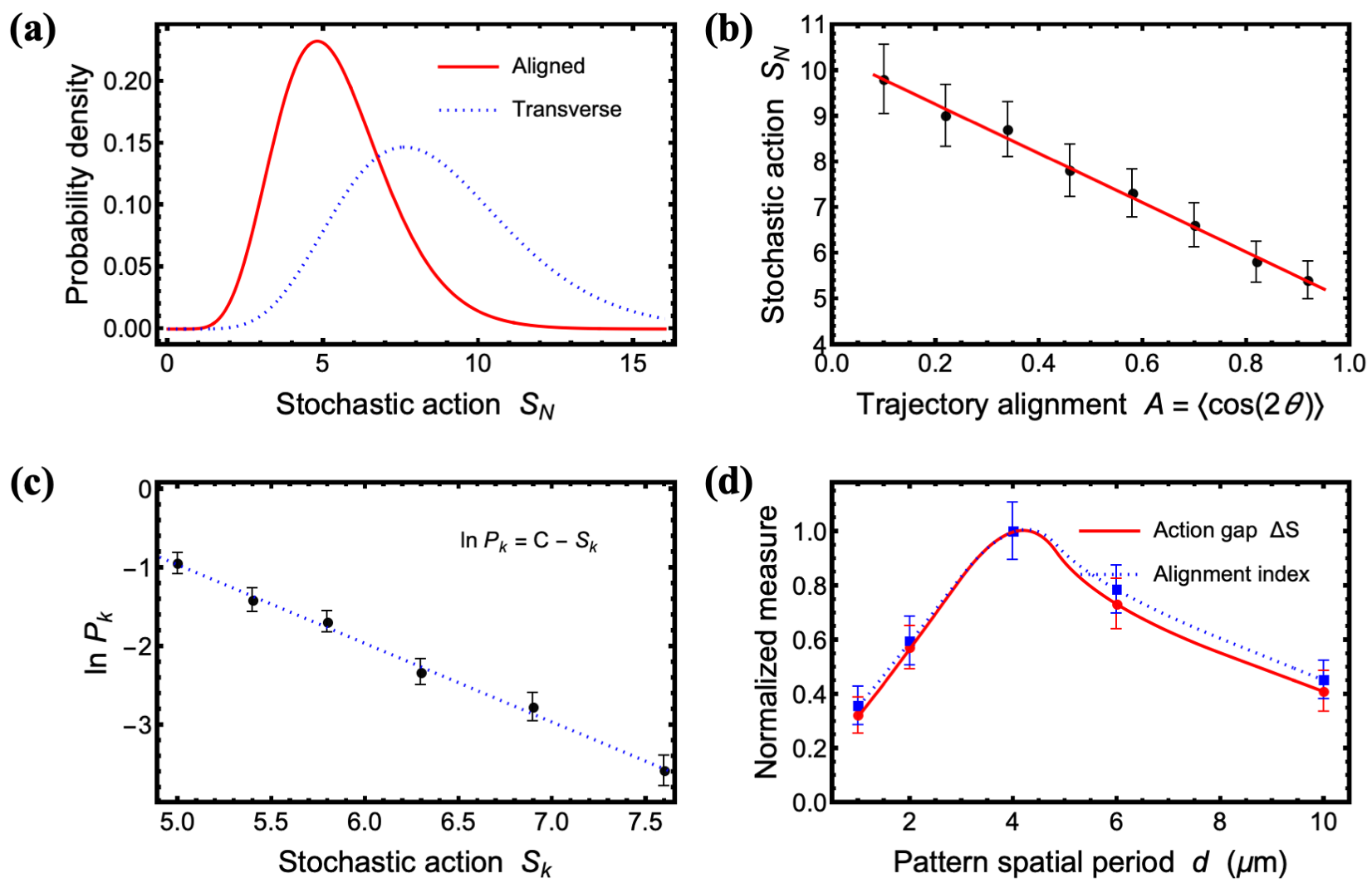}
    \caption{Stochastic 
 action analysis of axonal growth on periodic micropatterns.
    (\textbf{a}) {Model prediction for the probability density distributions of the discrete stochastic action }$S_N${ for predominantly aligned and transverse trajectory classes.}
    (\textbf{b}) {Experimental mean stochastic action }$S_N${ as a function of the trajectory alignment parameter }$A=-\langle\cos(2\theta)\rangle${. The points represent experimental data, error bars indicate uncertainties in the mean, and the red continuous curve is a linear fit to the data.}
    (\textbf{c})~{Experimental values of }$\ln P_k${ as a function of the mean stochastic action }$S_k${ for }$K=10${ trajectory classes defined using equal-width action intervals. The points represent experimental data, error bars indicate statistical uncertainties, and the blue dotted line is a fit to the data.}
    (\textbf{d})~{Experimental normalized action difference }$\Delta S=\langle S_N\rangle_{\mathrm{misaligned}}-\langle S_N\rangle_{\mathrm{aligned}}${ and normalized axonal alignment as functions of micropattern spatial period }$d${. Here, misaligned refers to trajectories with comparatively low values of the alignment parameter }$A${, indicating substantial deviation from the preferred directions parallel to the micropatterns, whereas aligned trajectories have larger values of }$A${ and remain predominantly oriented along the pattern direction. Error bars indicate uncertainties from the corresponding trajectory ensembles. The red continuous curve represents the normalized action difference, and the blue dotted curve represents the normalized alignment measure.}}
    \label{fig4}
\end{figure*}

{The analyses shown in Figure~}\ref{fig4}{ were performed using previously published experimental data from }$N=1840${ growth cones obtained in our previous studies of cortical neurons growing on micropatterned PDMS substrates }
\cite{Yurchenko2019,Vensi2019,Yurchenko2021,Descoteaux2022,Staii2023,Staii2025}.
{The stochastic-action analysis presented here is new and was performed using
the experimentally determined Langevin and Fokker--Planck parameters associated
with these measurements. The data provenance, parameter values and ranges,
sample sizes, and statistical analysis of the predicted action--probability
relationship, including the test of the expected slope }$-1${, are provided in
Appendix~}\ref{app:fig4_analysis}{.} 

Figure~\ref{fig4}a compares the distributions of stochastic action for trajectories classified as predominantly aligned or transversely directed. The probability density plotted in this panel is the normalized distribution \(p(S_N)\) of the discrete stochastic action values measured over the ensemble of trajectories in each class. Thus, \(p(S_N)\,\mathrm{d}S_N\) represents the fraction (probability) of trajectories in a given class with stochastic action between \(S_N\) and \(S_N+\mathrm{d}S_N\), with the total area under each distribution normalized to unity. This probability density should be distinguished from the probability assigned to an individual trajectory in Equation~\eqref{eq:path_probability}. The smooth curves in Figure~\ref{fig4}a represent probability density fits to the corresponding action distributions. A shift of the aligned distribution toward smaller \(S_N\) reflects the central prediction of the stochastic action formulation: trajectories that remain closer to the deterministic guidance field require smaller cumulative stochastic deviations and therefore accumulate lower action.

Figure~\ref{fig4}b examines this relationship directly by plotting the mean
stochastic action as a function of the trajectory alignment parameter \(A\),
with trajectories grouped into intervals of \(A\). The error bars represent
the uncertainty in the mean action within each group, and the red continuous
line is a linear regression used to characterize the dependence of \(S_N\) on
alignment. {Figure~}\ref{fig4}{c tests the predicted exponential
relationship between trajectory-class probability} \(P_k\) {and stochastic
action} \(S_k\) {at a coarse-grained level by grouping the measured
trajectories into classes} \(k\).
{The }$N_{\mathrm{tot}}=1840${ growth cone trajectories were divided into }
$K=10${ trajectory classes according to their stochastic action using
equal-width action intervals. This choice provides a sufficient number of
trajectories in each class while retaining adequate resolution of the action
distribution.}
{The probability of each of the} \mbox{$K=10$} {classes is estimated from its
observed frequency as} \(P_k=N_k/N_{\mathrm{tot}}\), {while} \(S_k\)
{is determined from the mean stochastic action of the trajectories belonging
to that class. Thus, each point in Figure~}\ref{fig4}{c is constructed from
two quantities determined from the same experimental trajectory class: its
observed fraction of the total trajectories and its corresponding mean
stochastic action. The occupancy of the 10 action classes and the corresponding
regression analysis are reported in Appendix~}\ref{app:fig4_analysis}{.}
{The blue dotted line in Figure~}\ref{fig4}{c is a linear fit to the
experimental class data. The stochastic-action relation predicts a slope of}
\(-1\) {for the dependence of} \(\ln P_k\) {on} \(S_k\) {when the
action intervals are sufficiently narrow and differences in their trajectory
space measure do not dominate the class probabilities. The agreement between
the fitted slope and this predicted value shows that the experimentally
measured class frequencies are consistent with the exponential action
probability relation predicted by the stochastic-action description
(details are provided in Appendix~}\ref{app:fig4_analysis}{).}

{It is important to distinguish the likelihood assigned to an individual
trajectory from the probability of a trajectory class. The stochastic weight }
$\exp[-S_N]${ applies to individual discretized trajectories, whereas the
class probability }$P_k${ represents the combined contribution of all
trajectories belonging to class }$k${. Consequently, }$P_k${ is not in
general equal to }$\exp[-S_k]${. The comparison in Figure~}\ref{fig4}c
{ should therefore be interpreted as a coarse grained test of the predicted
exponential action dependence, valid when the action distribution within each
class is sufficiently narrow and differences in trajectory space measure
between classes are small.}

Finally, the dependence of the stochastic action landscape on substrate geometry can be examined by comparing different micropattern spatial periods. Trajectories were classified according to the alignment parameter
$A=-\langle\cos(2\theta)\rangle$, with misaligned trajectories characterized
by lower values of $A$ and larger deviations from the substrate-preferred
directions, whereas aligned trajectories have higher values of $A$ and remain
predominantly oriented along the parallel micropatterns. For each value of \(d\), the action separation between trajectory classes is defined as \(\Delta S(d)=\langle S_N\rangle_{\mathrm{misaligned}}-\langle S_N\rangle_{\mathrm{aligned}}\), while the corresponding alignment parameter is obtained from the ensemble average $A(d)=-\langle\cos(2\theta)\rangle$. \mbox{Figure~\ref{fig4}d compares} these two quantities after normalization to their respective maximum values. The red continuous curve represents the normalized action difference, and the blue dotted curve represents the normalized alignment measure. The similar dependence of the two quantities on \(d\) indicates that the substrate geometry producing the strongest directional alignment also produces the largest separation in stochastic action between aligned and misaligned trajectory ensembles.

\subsubsection{Mechanical Origin of the Guidance Landscape}

The effective guidance potential and stochastic action describe the trajectory statistics, but the preferred directions ultimately arise from the mechanical interaction between the axon, growth cone, and patterned substrate. Our previous experiments and mechanical modeling showed that bending of an axon over curved PDMS features provides one physical contribution to this directional preference~\cite{Descoteaux2022,Staii2025}.

For an axon growing over a locally curved ridge of radius $R$, the bending energy can be written~\cite{Descoteaux2022,Staii2025} as
\begin{equation}
E(\theta)
=
\frac{B}{2}\kappa^2(\theta,R),
\label{eq:bending_energy_general}
\end{equation}
where $B$ is the axonal bending stiffness. For the locally semicylindrical geometry considered in the model, the curvature is $\kappa(\theta,R)=|\cos\theta|/R$, giving
\begin{equation}
E(\theta)
=
\frac{B}{2R^2}\cos^2\theta.
\label{eq:bending_energy}
\end{equation}

The bending energy is minimized for orientations parallel to the ridge direction. An effective near-stationary angular distribution of the form~\cite{Descoteaux2022,Staii2025}
\begin{equation}
p(\theta)
\propto
\exp\left[-\frac{E(\theta)}{E_0}\right]
\label{eq:angular_boltzmann}
\end{equation}
therefore favors the same orientations observed experimentally in Figure~\ref{fig2}. Here, $E_0$ is an effective fluctuation scale associated with the active cellular dynamics and is not interpreted as a thermal energy.

The mechanical and stochastic descriptions consequently operate at complementary levels. Substrate geometry modifies the bending and traction response of the growing axon, these mechanical effects generate an effective deterministic guidance field, and the combination of this drift with cellular fluctuations determines the probability of complete trajectories. The resulting hierarchy can be summarized as: % Note: This equation was reformatted into multiple lines and the adjustwidth
\begin{equation}
\begin{gathered}
\text{substrate geometry}
\rightarrow
\begin{gathered}
\text{mechanical bending}\\
\text{and traction response}
\end{gathered}
\\
\rightarrow
\text{deterministic drift}
\rightarrow
\text{stochastic path action}
\\
\rightarrow
\text{most probable trajectory}.
\end{gathered}
\label{eq:hierarchy}
\end{equation}

The minimum action principle obtained here therefore does not imply that an individual axon minimizes a conservative mechanical action. Rather, it states that, for the stochastic dynamics described by the experimentally motivated Langevin and Fokker--Planck models, lower-action trajectories have larger path probabilities and the most probable trajectories are minimum action paths under the specified boundary conditions.

\subsection{Durotactic Growth at a PDMS Stiffness Interface}
\label{sec:durotaxis_action}

The second {variational description} considered in this work concerns axonal
growth across a mechanical interface separating two PDMS regions with different
stiffnesses. As described in Section~\ref{sec:methods} and illustrated in
Figure~\ref{fig1}b, the substrate contains a stiffer region with Young's modulus $E_{s,1}=57\pm6$~kPa
 and a softer region with \mbox{$E_{s,2}=8\pm2$~kPa}.
 In contrast to the periodic micropatterns considered
above, these substrates do not impose a directional surface geometry. Instead,
the guidance cue arises from the dependence of growth cone adhesion and traction
on substrate stiffness. Growth cones respond mechanically to substrate rigidity through adhesion, actomyosin contractility, and molecular-clutch coupling between the cytoskeleton
and the extracellular environment~\cite{ChanOdde2008,Koch2012,Sens2013,
Bangasser2013}. These mechanisms can produce a nonlinear relation between
substrate stiffness and the traction generated by the growth cone. A mechanical
description of axonal durotaxis was developed in Refs.~\cite{Oliveri2021,Kassianides2025}, and incorporates the
traction response into the force acting on the growth cone and the
resulting deformation of the growing axon.
This framework provides the starting point for the {Fermat-like stationary action description} used here.

\subsubsection{Mechanical Basis of Stiffness-Dependent Guidance}

Let $E_s(\mathbf r)$ denote the local Young's modulus of the substrate and
$T(E_s)$ the corresponding traction generated by the growth cone. For a growth
cone of radius $R$, adhesion-site density $\rho$, and angular extent
$2\varphi_0$, integration of the local traction over the finite growth cone gives,
to leading order in the spatial variation of substrate stiffness
\cite{Oliveri2021,Kassianides2025},
\begin{equation}
\begin{aligned}
\mathbf F
&=\alpha_g T(E_s)\mathbf t
+\beta_g\frac{dT}{dE_s}
\\
&\qquad\times\mathbf P(\varphi_0)\nabla E_s+\cdots,
\end{aligned}
\label{eq:duro_force}
\end{equation}
where $\mathbf t$ is the local tangent to the axon,
$\alpha_g=2R\rho\sin\varphi_0$, $\beta_g=R^2\rho\varphi_0$, and
$\mathbf P(\varphi_0)$ is a tensor determined by the growth cone geometry. The
first term in Equation~\eqref{eq:duro_force} represents the traction acting
predominantly along the direction of growth, whereas the second term generates
a transverse force when the growth cone encounters a spatial change in
substrate stiffness.

For an approximately symmetric growth cone, the leading geometric tensor may
be approximated by the identity, $\mathbf P\simeq\mathbf I$. The
stiffness-dependent transverse force then takes the simpler form
\cite{Kassianides2025}
\begin{equation}
\mathbf F_{\mathrm{duro}}
=
\beta_g
\frac{dT}{dE_s}\nabla E_s
=
\beta_g\nabla T(E_s).
\label{eq:duro_gradient_force}
\end{equation}

Equation~\eqref{eq:duro_gradient_force} shows that the direction of mechanical
deflection is determined by the traction difference sensed across the growth
cone. Importantly, the sign of the response is controlled by the derivative of
the traction with respect to stiffness and not by stiffness alone. Thus, an
axon does not necessarily turn toward the mechanically stiffer region. Its
direction depends on whether the stiffer substrate produces larger or smaller
growth cone traction within the relevant range of $E_s$
\cite{Oliveri2021,Kassianides2025}.

For the two-region PDMS substrate used here, $E_s$ is approximately constant
away from the interface and changes over a narrow transition region (Figure~\ref{fig1}b). We therefore associate each side of the interface with a traction
$T_i=T(E_{s,i})$, where $T_1=T(E_{s,1})$ and $T_2=T(E_{s,2})$. The finite
growth cone samples the mechanical transition as it approaches the boundary,
producing the transverse force responsible for changes in the direction of
axonal growth.

\subsubsection{Fermat-Like Action and Effective Axonal Refractive Index}

The predicted refraction of axonal trajectories at a stiffness interface
suggests a reduced description analogous to Fermat's principle in geometrical
optics~\cite{BornWolf1999}. The complete durotactic dynamics is
active and dissipative and therefore cannot, in general, be represented by a
conservative optical ray model~\cite{Oliveri2021,Kassianides2025}. However, when mechanical relaxation of the
distal axon is rapid compared with growth and the local deformation remains
small, the trajectory can be represented by a spatial stationary-action
functional.

Let the axonal trajectory be a planar curve
$\Gamma:\mathbf r(s)$ parameterized by arclength $s$. We introduce the
Fermat-like functional
\begin{equation}
\mathcal A_{\mathrm F}[\Gamma]
=
\int_\Gamma
n_{\mathrm{ax}}[T(E_s(\mathbf r))]\,ds,
\label{eq:fermat_action}
\end{equation}
where $n_{\mathrm{ax}}$ is an effective axonal refractive index determined by
the local traction response~\cite{Oliveri2021,Kassianides2025}. The preferred path satisfies the stationary action condition
\begin{equation}
\delta\mathcal A_{\mathrm F}=0.
\label{eq:fermat_stationary}
\end{equation}

The effective index can be connected to the mechanical response of the growing
axon using the quasi-static beam description developed for axonal durotaxis
\cite{Oliveri2021,Kassianides2025}. In the distal mechanically active region,
the small-deflection beam equation is~\cite{Barber2022}
\begin{equation}
\mathcal B y''''-F_t y''=0,
\label{eq:beam_static}
\end{equation}
where $y$ is the deflection, $\mathcal B$ is the axonal bending stiffness, and $F_t$ is the tangential
force generated by the growth cone. For an active distal segment of length
$L$, the curvature generated by a transverse force $F_n$ is
\cite{Oliveri2021,Kassianides2025}
\begin{equation}
\kappa
=
\frac{F_n}{\sqrt{\mathcal B F_t}}
\tanh\left(
L\sqrt{\frac{F_t}{\mathcal B}}
\right).
\label{eq:beam_curvature}
\end{equation}
Using the leading growth cone forces
$F_t\simeq\alpha_gT(E_s)$ and
$F_n\simeq\beta_g(dT/dE_s)\,
\mathbf n\cdot\nabla E_s$, where $\mathbf n$ is the normal to the axonal
trajectory, Equation~\eqref{eq:beam_curvature} becomes
\begin{equation}
\kappa
=
\frac{\beta_g}{\sqrt{\mathcal B\alpha_gT}}
\tanh\left(
L\sqrt{\frac{\alpha_gT}{\mathcal B}}
\right)
\frac{dT}{dE_s}\,
\mathbf n\cdot\nabla E_s.
\label{eq:kappa_T}
\end{equation}
A derivation of Equations~\eqref{eq:beam_static} and
\eqref{eq:beam_curvature}, together with the resulting
stiffness-dependent curvature relation, is provided in
Appendix~\ref{app:beam_derivation}.

For a Fermat-like trajectory, variation of
Equation~\eqref{eq:fermat_action} gives the corresponding curvature relation
$\kappa=\mathbf n\cdot\nabla\ln n_{\mathrm{ax}}$ (see Appendix~\ref{app:fermat_curvature}). Comparison with
Equation~\eqref{eq:kappa_T} therefore gives
\begin{equation}
\frac{d\ln n_{\mathrm{ax}}}{dT}
=
\frac{\beta_g}{\sqrt{\mathcal B\alpha_gT}}
\tanh\left(
L\sqrt{\frac{\alpha_gT}{\mathcal B}}
\right).
\label{eq:index_diff}
\end{equation}

Integration of Equation~\eqref{eq:index_diff} gives the effective axonal
refractive index
\begin{equation}
n_{\mathrm{ax}}(T)
=
n_0
\left[
\cosh\left(
L\sqrt{\frac{\alpha_gT}{\mathcal B}}
\right)
\right]^{2\beta_g/(\alpha_gL)},
\label{eq:effective_index}
\end{equation}
where $n_0$ is an arbitrary positive normalization constant. This relation is
obtained here by combining the growth cone traction law with the quasi-static
beam mechanics. It is not an additional constitutive assumption. Since
$dn_{\mathrm{ax}}/dT>0$, a region producing larger growth cone traction has a
larger effective axonal refractive index within this reduced description.
Equation~\eqref{eq:effective_index} {determines the effective index as a
function of growth cone traction, but does not by itself specify the relation
between }$n_{\mathrm{ax}}${ and substrate Young's modulus, which additionally
requires knowledge of the traction--stiffness relation }$T(E_s)$. {Direct growth cone traction measurements were not performed on the two PDMS
regions in the present study, and establishing the corresponding traction--stiffness
relation experimentally will be the subject of future work.}

\subsubsection{Refraction at the Two-Region PDMS Interface}
\label{sec:pdms_refraction}

\begin{figure*}[t]
\centering
    \includegraphics[width=0.85\textwidth]{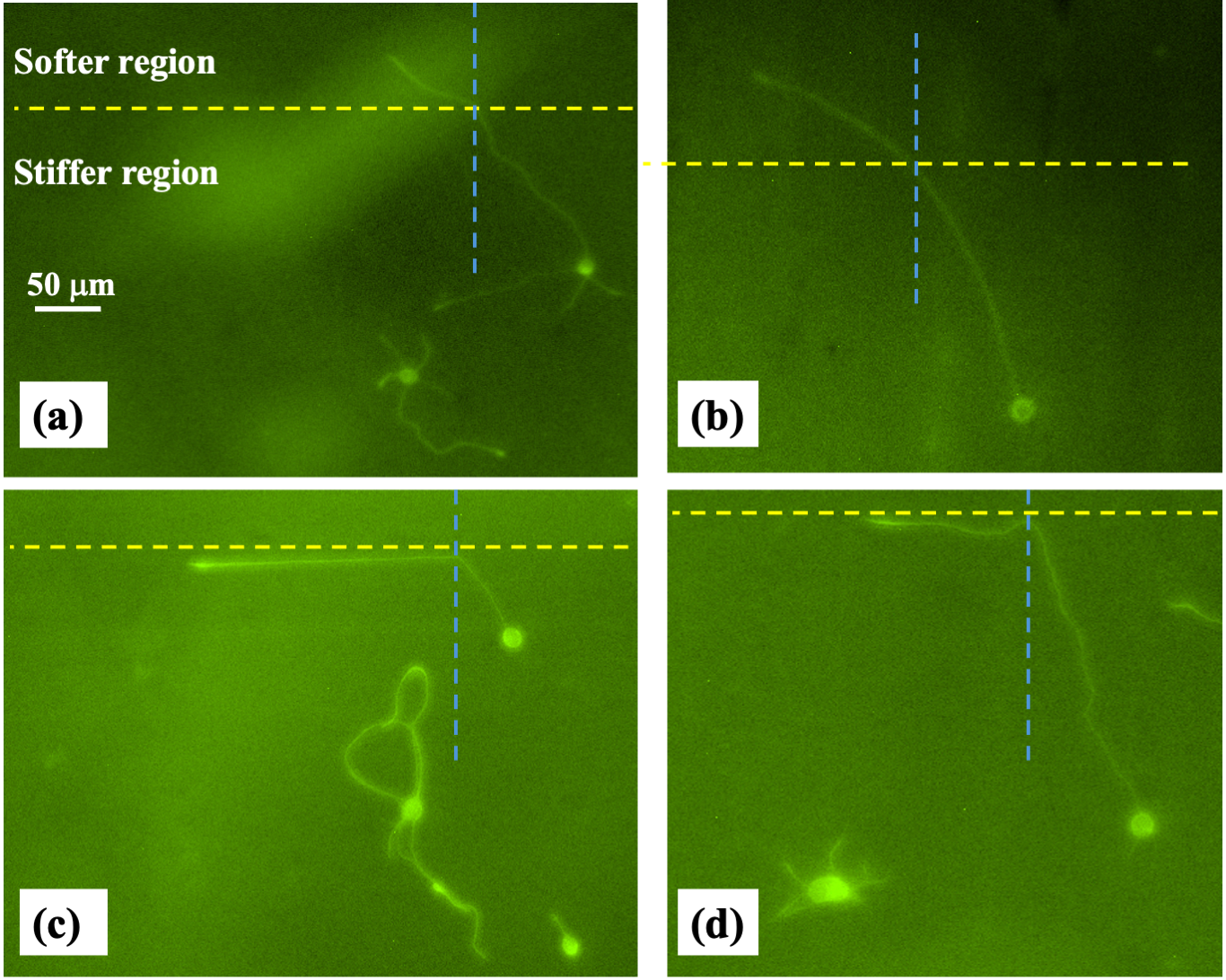}
    \caption{Axonal 
 refraction at the interface between PDMS regions
    of different stiffness. The panels show representative fluorescence images of axons
    approaching the stiffness boundary from the stiffer PDMS region
    ($E_{s,1}=57\pm6$~kPa)
    and crossing into the softer region ($E_{s,2}=8\pm2$~kPa). The yellow dotted
    line indicates the boundary between the two regions, and the blue dotted line
    indicates the normal to the interface used to define the incident angle
    $\alpha_1$ and transmitted angle $\alpha_2$. Representative incident angles are
    $\alpha_1\simeq29^\circ$ in (\textbf{a}) and
    $\alpha_1\simeq23^\circ$ in (\textbf{b}). The trajectories in
    (\textbf{c}),(\textbf{d}) {show examples of growth consistent with
    incidence near the critical angle behavior predicted by the reduced model,
    with an approximate limiting angle on the order of }
    $\alpha_c\simeq38^\circ${ suggested by these representative observations.}
    In all four examples, the transmitted axonal trajectory forms a larger angle
    with the interface normal than the incident trajectory, indicating refraction
    away from the normal after entering the softer PDMS region. The scale bar shown
    in (\textbf{a}) applies to all images.}
    \label{fig5}
\end{figure*}

For the PDMS substrate shown in Figure~\ref{fig1}b, the two regions are
approximately homogeneous away from their common boundary. The corresponding
effective axonal refractive indices are therefore
\begin{equation}
n_1=n_{\mathrm{ax}}[T(E_{s,1})],
\qquad
n_2=n_{\mathrm{ax}}[T(E_{s,2})],
\label{eq:n12}
\end{equation}
with $E_{s,1}=57\pm6$~kPa for the stiffer region and  $E_{s,2}=8\pm2$~kPa for the softer region.

Consider an axon approaching
the interface from region 1 at an incident angle $\alpha_1$ and crossing into
region 2 at a transmitted angle $\alpha_2$. Both angles are measured relative
to the normal to the stiffness boundary (Figure~\ref{fig5}). For fixed initial and final positions, the Fermat-like action of a transmitted
trajectory can be written as the sum of the contributions accumulated in the
two regions,
\begin{equation}
\mathcal A_{\mathrm F}
=
n_1\ell_1+n_2\ell_2,
\label{eq:two_region_action}
\end{equation}
where $\ell_1$ and $\ell_2$ are the path lengths in the first and second
regions, respectively. Varying the position at which the trajectory crosses
the interface and imposing stationarity of
Equation~\eqref{eq:two_region_action} gives the Snell-like relation
\begin{equation}
n_1\sin\alpha_1
=
n_2\sin\alpha_2.
\label{eq:axon_snell}
\end{equation}

Equation~\eqref{eq:axon_snell} is the central prediction of the reduced
{stationary action} description for refraction at the two-region PDMS
interface. Combining this relation with the effective index in
Equation~\eqref{eq:effective_index} gives
\begin{equation}
\begin{aligned}
\frac{\sin\alpha_1}{\sin\alpha_2}
&=\frac{n_2}{n_1}
\\
&=\left[\frac{\cosh\!\left(L\sqrt{\alpha_gT_2/\mathcal B}\right)}
{\cosh\!\left(L\sqrt{\alpha_gT_1/\mathcal B}\right)}
\right]^{2\beta_g/(\alpha_gL)},
\end{aligned}
\label{eq:snell_traction}
\end{equation}
{where }$T_1=T(E_{s,1})${ and } $T_2=T(E_{s,2})$ { denote the growth cone tractions associated with regions 1 and 2, respectively. As discussed above, quantitative evaluation of these tractions from the substrate stiffnesses requires the corresponding traction--stiffness relation }$T(E_s)${. Although quantitative molecular-clutch models can provide such a relation,
their evaluation for the present neuronal cultures requires clutch and
cytoskeletal parameters that were not independently measured in these
experiments}~\cite{ChanOdde2008,Sens2013,Bangasser2013,Kassianides2025}. {Therefore, assigning numerical values to }$T_1${ and }$T_2${ from
the measured substrate stiffnesses would require additional assumptions that
cannot be independently validated from the present experiments.} {Within the present framework, the
change in axonal direction is therefore governed by the traction contrast
between the two substrate regions rather than directly by the ratio of their
Young's moduli. Equation~}\eqref{eq:snell_traction}{ provides the resulting
traction-dependent prediction for the ratio of the refractive index, which can be
evaluated quantitatively once }$T(E_s)${ is established experimentally.}

Equation~\eqref{eq:axon_snell} predicts two qualitatively different
refraction regimes. {If }$T_2>T_1${, then
Equation~}\eqref{eq:effective_index}{ gives }$n_2>n_1${, and therefore }
$\alpha_2<\alpha_1${. In this case, the transmitted axon bends toward the
interface normal. Conversely, if }$T_2<T_1${, then }$n_2<n_1${ and }
$\alpha_2>\alpha_1${, corresponding to refraction away from the normal.
These should be interpreted as predictions based on the traction
contrast rather than as assignments determined directly from substrate
stiffness.} This distinction is important because molecular clutch models
can produce a nonlinear and non-monotonic relationship between substrate
stiffness and cellular traction
\cite{ChanOdde2008,Sens2013,Bangasser2013,Kassianides2025}.
{Consequently, a larger substrate Young's modulus does not by itself imply
a larger growth cone traction or a larger effective axonal refractive index.}

Representative experimental examples of axonal refraction at the two-region
PDMS interface are shown in Figure~\ref{fig5}. In all four examples, the axon
approaches the boundary from the stiffer region,
$E_{s,1}=57\pm6$~kPa and enters the softer region,
$E_{s,2}=8\pm2$~kPa. The stiffness boundary is indicated by a yellow dotted
line, while the blue dotted line indicates the normal used to define
$\alpha_1$ and $\alpha_2$. The experimentally observed direction of refraction
provides information about the effective index contrast. Since
$\alpha_2>\alpha_1$ for the examples in Figure~\ref{fig5},
Equation~\eqref{eq:axon_snell} requires
$n_1>n_2$ and therefore {corresponds to} a decrease in the effective axonal
refractive index as the axon crosses from the stiffer PDMS region into the
softer region. Because the effective index in
Equation~\eqref{eq:effective_index} is an increasing function of traction,
{this observation also implies }$T_1>T_2${ within the reduced model.}
The measured trajectories are therefore consistent with a decrease in both
growth cone traction and effective axonal refractive index as the axon crosses
from the stiffer PDMS region into the softer region.
{This inference applies to the representative trajectories observed at
these two substrate stiffnesses} and does not require the traction--stiffness
relation to be monotonic over its entire range
\cite{ChanOdde2008,Sens2013,Bangasser2013,Kassianides2025}.

{For transmission from region 1, which is inferred within the reduced model
to have the higher effective index, into region 2, with }$n_1>n_2${, the
Snell-like relation predicts a critical incident angle}:
\begingroup
\setlength{\abovedisplayskip}{5pt}
\setlength{\belowdisplayskip}{5pt}
\setlength{\abovedisplayshortskip}{3pt}
\setlength{\belowdisplayshortskip}{3pt}
\begin{equation}
\sin\alpha_c
=
\frac{n_2}{n_1},
\qquad
n_1>n_2.
\label{eq:critical_angle}
\end{equation}
\endgroup
As $\alpha_1$ approaches $\alpha_c$ from below,
Equation~\eqref{eq:axon_snell} predicts that the transmitted angle approaches
$\alpha_2=90^\circ$, corresponding to an axonal trajectory approximately
parallel to the stiffness boundary. For $\alpha_1>\alpha_c$, the idealized
Fermat-like model has no transmitted solution. This critical angle condition
should be interpreted as a limiting prediction of the reduced description
rather than as an exact optical law for the active biological system.

{The representative trajectories shown in Figure~}\ref{fig5}{ provide
examples of axonal refraction at the interface between the stiffer and softer
PDMS regions. In these examples, the transmitted trajectories bend away from
the interface normal, consistent with the qualitative behavior predicted by
the Fermat-like model. Some representative trajectories become approximately
parallel to the interface at incident angles on the order of }
$38^\circ${. However, these observations are not used to determine a
critical angle quantitatively. No population-level statistical analysis or fit
of transmission probability as a function of incident angle was performed in
the present study. Consequently, the value }$38^\circ${ should be regarded
only as an illustrative angle observed in the representative trajectories and
not as a statistically established critical angle.}

\section{Discussion}
\label{sec:discussion}

The results presented here show that two experimentally accessible forms of
axonal guidance can be formulated in terms of well-defined variational
principles. The significance of this result is not that growing axons obey a
single universal minimum action law analogous to conservative mechanics.
Rather, different levels of the axonal dynamics admit different action
functionals whose physical meanings follow from the corresponding stochastic
or mechanical descriptions. This distinction is particularly important for
neuronal growth, where active cytoskeletal processes, adhesion, traction,
feedback, and stochastic fluctuations all contribute to trajectory selection.

For neurons growing on periodic parallel micropatterns, the minimum action
principle follows directly from the experimentally calibrated Langevin and
Fokker--Planck description. The stochastic action measures the cumulative
deviation of a trajectory from the deterministic guidance drift, weighted by
the strength of the fluctuations. Lower action trajectories therefore require
smaller stochastic departures from the substrate-guided dynamics and have
larger path probabilities. This provides a precise statistical interpretation
of minimum action for axonal growth and places the experimentally motivated
model within the broader Onsager--Machlup and Freidlin--Wentzell framework for
stochastic dynamics~\cite{OnsagerMachlup1953,FreidlinWentzell2012}. It is also
consistent with path-information approaches in which trajectories acquire
exponentially decreasing statistical weights with increasing action
\cite{Wang2005,Wang2009}.

An important feature of this formulation is that the action does not require
the introduction of an additional phenomenological guidance mechanism. The
same substrate and feedback terms that describe the measured growth cone
dynamics determine the path action. Consequently, quantities already obtained
from trajectory measurements, including drift, diffusion, pattern spacing,
and angular alignment, can be used to evaluate the relative statistical cost
of different axonal paths. The analyses illustrated in Figure~\ref{fig4}
therefore suggest direct experimental tests based on the separation of action
distributions for aligned and misaligned trajectories, the dependence of mean
action on directional alignment, and the exponential relation between
trajectory probability and stochastic action. A particularly useful
prediction is that substrate geometries producing stronger directional
guidance should also produce a larger action separation between preferred and
nonpreferred trajectory classes.

The mechanical bending analysis provides a complementary interpretation of
this stochastic result. The patterned substrate modifies the mechanical
response of the growing axon and growth cone, which contributes to the
deterministic guidance field. The stochastic action then operates at the level
of complete trajectories and quantifies fluctuations around that field. The
mechanical energy associated with local bending and the stochastic path action
should therefore not be identified with one another. Instead, they describe
successive levels of the same guidance process: substrate geometry modifies
the local mechanical response, the mechanical response contributes to the
effective drift, and the combination of drift and fluctuations determines the
probability of a complete growth trajectory.

The stiffness-interface problem leads to a different variational description.
Starting from the stiffness-dependent growth cone traction and the
quasi-static bending response of the distal axon
\cite{Oliveri2021,Kassianides2025}, we derived an effective axonal refractive
index that depends on traction. This result allows axonal turning at a
mechanical interface to be represented by a Fermat-like stationary action and
leads to a Snell-like relation between the incident and transmitted growth
angles. The effective index is therefore not introduced as an empirical
optical analogy, but is obtained by matching the curvature generated by the
durotactic beam mechanics to the curvature generated by the spatial
stationary-action functional.

The experimental trajectories in Figure~\ref{fig5} provide a direct example
of this interpretation. The axons cross from the stiffer PDMS region,
$E_{s,1}=57\pm6$~kPa, into the softer region,
$E_{s,2}=8\pm2$~kPa, and bend away from the interface normal. Within the
reduced Fermat-like description, the observed relation
$\alpha_2>\alpha_1$ requires $n_1>n_2$. Since the derived effective index is
an increasing function of growth cone traction, the same observation implies
$T_1>T_2$ for these two substrate stiffnesses. {The representative trajectories are consistent with the critical angle
behavior predicted by the reduced model, with incident angles on the order of}
$38^\circ$ {observed in examples where the transmitted trajectories become
approximately parallel to the interface.} These relations connect directly measurable trajectory
angles to otherwise less accessible mechanical quantities such as the
effective traction contrast between the two regions.

The two {variational approaches} are therefore complementary but physically
distinct (Table~\ref{Table1}). On the periodic micropatterns, the action is a stochastic functional
defined in time, and minimizing it identifies the most probable trajectory
within an ensemble of noisy paths. At the stiffness interface, the Fermat-like
action is a reduced spatial functional, and its stationary trajectories
describe the geometry of deterministic refraction in a quasi-static limit.
The first is fundamentally probabilistic, whereas the second is a reduced
mechanical description. Their common element is the deterministic guidance
field generated by interactions between the cell and its substrate. Geometry
produces this field through contact guidance, bending, and feedback on the
periodic patterns, whereas stiffness-dependent traction and axonal deformation
produce it at the mechanical interface. Cellular fluctuations can, in
principle, be added to either problem, making the stochastic path description
the broader framework for active axonal dynamics.

\begin{table*}[t]
\caption{{Comparison 
 of the two variational descriptions used in this work.}}
\label{tab:action_comparison}
\centering
\footnotesize
\setlength{\tabcolsep}{3pt}
\renewcommand{\arraystretch}{1.15}

\resizebox{\textwidth}{!}{%
\begin{tabular}{lll}
\toprule
&
\textbf{Micropattern Guidance} &
\textbf{Stiffness-Interface Guidance} \\
\midrule

State variable &
Transverse growth cone position $x(t)$, or more generally $\mathbf r(t)$ &
Spatial axonal path $\mathbf r(s)$ \\

Parameterization &
Time $t$ &
Arclength $s$ \\

Underlying model &
Langevin/Fokker--Planck stochastic dynamics &
Quasi-static beam mechanics and stiffness-dependent traction \\

\midrule

Variational quantity &
Discrete path likelihood / Freidlin--Wentzell rate functional &
Fermat-like spatial action \\

Variational condition &
Minimum action for the most probable path under specified endpoints &
Stationary spatial action \\

Physical interpretation &
Statistical cost of deviations from deterministic drift &
Reduced geometrical description of mechanical refraction \\

\midrule

Principal assumptions &
Markovian additive Gaussian noise; reduced transverse description &
Small deflection; quasi-static relaxation; reduced growth cone mechanics \\

Scope &
Noisy trajectory probabilities &
Deterministic refraction geometry \\

\bottomrule
\end{tabular}
}
\label{Table1}
\end{table*}

This comparison also clarifies the novelty of the present work. Previous
studies established stochastic models for neuronal growth on patterned
substrates~\cite{Yurchenko2019,Vensi2019,Descoteaux2022,Staii2023,Staii2025}
and mechanical models for axonal durotaxis and refraction
\cite{Oliveri2021,Kassianides2025}. Here, these descriptions are reformulated
at the trajectory level. For geometrical guidance, the experimentally
calibrated stochastic dynamics is converted into a path action that assigns
relative probabilities to complete growth trajectories. For stiffness
guidance, the traction law and beam mechanics are combined to obtain an
effective axonal refractive index and a corresponding stationary-action
description of refraction. The resulting framework therefore connects local
models of growth cone motion and mechanics to global constraints on axonal
paths.

The same conceptual approach may be useful beyond neuronal growth. Many forms
of cell migration are modeled as a combination of systematic drift generated
by chemical, mechanical, or geometrical cues and fluctuations arising from
active intracellular processes~\cite{Alert2020,amselem2012stochastic,
li2011dicty}. Whenever these quantities can be measured or inferred, a
stochastic path action provides a way to compare complete migration
trajectories rather than only local velocities or ensemble averages.
Similarly, mechanically induced turning at boundaries between heterogeneous
materials can potentially be analyzed using reduced spatial actions when a
clear separation exists between rapid mechanical relaxation and slower cell
motion. The framework developed here therefore suggests a route for connecting
stochastic cell motility, mechanosensing, and trajectory selection within a
common path-based language.

These results also have implications for the design of biomimetic growth
environments. Microfabricated and mechanically patterned substrates are
increasingly used to control neuronal orientation, elongation, and connectivity
\cite{Teixeira2020,Descoteaux2022,Spedden2012Elasticity,Koch2012,
CastroDominguez2023,Pfister2011}. The present formulation suggests that such
substrates can be characterized not only by average alignment or growth speed,
but also by how strongly they separate desired and undesired trajectory
families in action or by how mechanical interfaces redirect growing axons.
This provides a potential quantitative basis for inverse design of guidance
substrates. Pattern geometry could be selected to maximize the stochastic
preference for a desired direction, while stiffness contrasts and interfaces
could be designed to produce controlled turning or refraction. Such principles
are relevant to engineered neuronal networks, nerve-guidance scaffolds, and
strategies intended to promote directed regeneration and functional
reconnection~\cite{Pfister2011,Holland2015,Bayly2014,Mahar2018,
Teixeira2020}.

{Related studies have connected substrate-imposed neuronal anisotropy with
network-scale structure and dynamics. Montalà-Flaquer et al. showed that
topographically patterned neuronal cultures develop activity propagation and
functional connectivity that reflect the imposed substrate geometry }
\cite{MontalaFlaquer2022}{, while Houben et al. used simulations combining
anisotropic connectivity with neuronal dynamics to examine how structural
anisotropy shapes collective network behavior }
\cite{Houben2025}{. The trajectory-level framework developed here is
complementary and could provide experimentally constrained inputs for future
multiscale models linking axonal path selection to network structure and
activity. Future studies combining trajectory measurements with calcium sensitive
fluorescence imaging could relate substrate-guided axonal path selection
and morphological changes to spontaneous neuronal activity and functional
connectivity.} 

Several limitations define the range of validity of the present description.
For the micropatterned substrates, the simplest stochastic action assumes
Markovian dynamics with additive Gaussian fluctuations.
{Gaussianity, temporal independence, stationarity, and strictly additive
noise were not independently assessed for the present dataset and should
therefore be understood as modeling assumptions inherited from the
coarse-grained Langevin/Fokker--Planck description. Future trajectory datasets
can test these assumptions directly through increment distributions and
temporal correlations.}
Memory effects, multiplicative noise, strongly non-Gaussian statistics,
branching, pausing, retraction, and interactions among neighboring cells would
require an enlarged state space or a generalized path functional. In
particular, internal variables describing cytoskeletal organization, adhesion
dynamics, or growth cone polarity may introduce correlations that are not
fully represented by a position-only Langevin model. These extensions are
conceptually compatible with the present framework but would require
additional experimentally determined variables and parameters. 

In addition, the Fermat-like description relies on small deformations, rapid
mechanical relaxation, and a reduced representation of the finite growth cone
and distal axon. {The effective refractive index is determined within the
model by growth cone traction, while quantitative connection to substrate
stiffness requires the traction--stiffness relation }$T(E_s)${. Direct
growth cone traction measurements were not performed for the two PDMS regions
used here, and experimental determination of this relation will be addressed
in future work.}
The physical transition between the two PDMS regions has a finite width (\mbox{Figures \ref{fig1}b and \ref{figA1}}), and the growth cone samples both mechanical environments while crossing the
interface. Moreover, the complete durotactic dynamics is active, dissipative,
and not generally reciprocal
\cite{Kassianides2025}.
The Snell-like and critical angle relations should therefore be viewed as
leading predictions for transmitted trajectories rather than exact optical
laws. {The representative trajectories are consistent with the predicted
critical angle behavior, but establishing a critical angle quantitatively
would require a population-level analysis of transmission as a function of
incident angle. Deviations from these predictions provide useful mechanistic information by revealing finite growth cone sensing, nonlinear traction responses, or additional internal mechanical degrees of freedom.} 

Building on these results, the principal advantage of the variational
description is that it converts experimentally measurable local dynamics into
quantitative statements about complete paths. For stochastic guidance, this
means ranking trajectories by probability through their action. For mechanical
refraction, it means relating trajectory geometry to substrate-dependent
traction through an effective index. This connection between local cellular
mechanics, stochastic fluctuations, and global path selection provides a
quantitative framework for interpreting neuronal guidance and for developing
engineered biomimetic environments in which axonal trajectories can be quantitatively
controlled.

\section{Conclusions}
\label{sec:conclusions}

{We have developed two complementary variational descriptions of axonal
growth on engineered PDMS substrates}. For periodic parallel micropatterns,
experimentally motivated Langevin and Fokker--Planck dynamics lead directly to
a stochastic path action in which lower action trajectories have larger
probabilities and the most probable paths minimize the cumulative deviation
from the deterministic guidance drift. For a boundary between regions of
different stiffness, stiffness-dependent growth cone traction and quasi-static
axonal bending lead to an effective axonal refractive index and {a Fermat-like stationary action that predicts a refraction relation and a
model dependent critical angle condition.}

These two actions have different physical meanings, but together they show how
substrate geometry and mechanics can be translated into quantitative
constraints on complete axonal trajectories. The framework links measurable
properties including drift, diffusion, stiffness, traction, alignment, and
refraction angles and provides direct experimental tests of path selection.
More broadly, it offers a path-based approach for modeling active cellular
guidance and suggests quantitative principles for designing biomimetic
substrates and guidance environments for controlled neuronal growth and
regeneration.

\section*{Acknowledgments}
This research was funded by a Tufts University Faculty Research Award (Cristian Staii).
\section*{Data Availability}
The original contributions presented in this study are included in the article. Further inquiries can be directed to the corresponding author.

\clearpage
\onecolumngrid
\begin{center}
{\bfseries\large Supplemental Material for Stationary Action Principles for Axonal Growth on Micropatterned Substrates}
\end{center}
\setcounter{section}{0}
\setcounter{subsection}{0}
\setcounter{equation}{0}
\renewcommand{\thesection}{\Alph{section}}
\renewcommand{\theequation}{\Alph{section}\arabic{equation}}
\makeatletter
\@addtoreset{equation}{section}
\makeatother
\section{Derivation of the Freidlin--Wentzell Rate Functional}
\label{app:fw_derivation}

Here we derive the continuum stochastic rate functional used in
Equation~\eqref{eq:micro_action} from the Langevin equation describing
the transverse motion of the growth cone. The starting point is the
one-dimensional stochastic differential equation
\begin{equation}
\dd x
=
b(x,t)\dd t
+
\sqrt{2D}\,\dd W_t,
\label{eq:app_sde}
\end{equation}
where $b(x,t)$ is the deterministic drift, $D$ is the effective
motility coefficient, and $W_t$ is a Wiener process. For the
micropatterned growth model considered in the main text, the drift is
generated by the effective guidance potential,
\begin{equation}
b(x,t)
=
-\frac{1}{\gamma}
\frac{\partial U_{\mathrm{eff}}(x,t)}{\partial x}.
\label{eq:app_gradient_drift}
\end{equation}
Here $\gamma$ is the effective damping coefficient. Because axonal
growth is an active nonequilibrium process, $D$ is interpreted as an
effective motility coefficient and need not satisfy an equilibrium
Einstein relation.

To obtain the probability associated with a complete trajectory,
divide the time interval $[t_i,t_f]$ into $N$ subintervals of duration
$\Delta t$, with
\begin{equation}
t_n=t_i+n\Delta t,
\qquad
n=0,\ldots,N,
\end{equation}
and define
\begin{equation}
x_n=x(t_n),
\qquad
\Delta x_n=x_{n+1}-x_n.
\end{equation}
Using an Euler--Maruyama discretization of
Equation~\eqref{eq:app_sde}, the increment over one time interval is
\begin{equation}
\Delta x_n
=
b(x_n,t_n)\Delta t
+
\sqrt{2D\Delta t}\,\xi_n,
\label{eq:app_increment}
\end{equation}
where the variables $\xi_n$ are independent Gaussian random variables
with zero mean and unit variance,
\begin{equation}
\langle \xi_n\rangle=0,
\qquad
\langle \xi_n\xi_m\rangle=\delta_{nm}.
\end{equation}
Solving Equation~\eqref{eq:app_increment} for $\xi_n$ gives
\begin{equation}
\xi_n
=
\frac{
\Delta x_n-b(x_n,t_n)\Delta t
}{
\sqrt{2D\Delta t}
}.
\label{eq:app_xi}
\end{equation}

The probability density for a particular value of $\xi_n$ is
\begin{equation}
p(\xi_n)
=
\frac{1}{\sqrt{2\pi}}
\exp\left(-\frac{\xi_n^2}{2}\right).
\end{equation}
Equation~\eqref{eq:app_xi} therefore gives the short-time transition
probability
\begin{equation}
P(x_{n+1}|x_n)
\simeq
\frac{1}{\sqrt{4\pi D\Delta t}}
\exp\left[
-\frac{
\left[
\Delta x_n-b(x_n,t_n)\Delta t
\right]^2
}{
4D\Delta t
}
\right].
\label{eq:app_transition_probability}
\end{equation}
For nonlinear drift, this expression should be understood as the
short-time transition probability associated with the discretized
stochastic dynamics.

Because Equation~\eqref{eq:app_sde} defines a Markov process, the
conditional probability of the discrete trajectory
\begin{equation}
\Gamma=\{x_0,x_1,\ldots,x_N\}
\end{equation}
is the product of the transition probabilities,
\begin{equation}
\mathcal P[\Gamma|x_0]
=
\prod_{n=0}^{N-1}
P(x_{n+1}|x_n).
\label{eq:app_path_product}
\end{equation}
Substitution of Equation~\eqref{eq:app_transition_probability} into
Equation~\eqref{eq:app_path_product} gives
\begin{align}
\mathcal P[\Gamma|x_0]
&\simeq
\left(4\pi D\Delta t\right)^{-N/2}
\nonumber\\
&\quad\times
\exp\left[
-\sum_{n=0}^{N-1}
\frac{
\left[
\Delta x_n-b(x_n,t_n)\Delta t
\right]^2
}{
4D\Delta t
}
\right].
\label{eq:app_discrete_path_probability}
\end{align}
The normalization factor is independent of the particular trajectory
when $D$, $\Delta t$, and $N$ are fixed. The trajectory-dependent
part of the probability can therefore be written as
\begin{equation}
\mathcal P[\Gamma|x_0]
\propto
\exp[-S_N(\Gamma)],
\label{eq:app_probability_action}
\end{equation}
where
\begin{equation}
S_N(\Gamma)
=
\sum_{n=0}^{N-1}
\frac{
\left[
\Delta x_n-b(x_n,t_n)\Delta t
\right]^2
}{
4D\Delta t
}.
\label{eq:app_discrete_action}
\end{equation}
We now consider a sufficiently regular candidate path $x(t)$. In the
continuum limit,
\begin{equation}
\Delta x_n
=
\dot{x}(t_n)\Delta t
+
\mathcal O(\Delta t^2).
\label{eq:app_continuum_increment}
\end{equation}
Consequently,
\begin{align}
\Delta x_n-b(x_n,t_n)\Delta t
&=
\left[
\dot{x}(t_n)-b(x_n,t_n)
\right]\Delta t
+
\mathcal O(\Delta t^2),
\end{align}
and each term in Equation~\eqref{eq:app_discrete_action} becomes
\begin{equation}
\frac{
\left[
\Delta x_n-b(x_n,t_n)\Delta t
\right]^2
}{
4D\Delta t
}
=
\frac{\Delta t}{4D}
\left[
\dot{x}(t_n)-b(x_n,t_n)
\right]^2
+
\mathcal O(\Delta t^2).
\label{eq:app_riemann_term}
\end{equation}
Thus, in the limit $\Delta t\rightarrow0$, the sum becomes the
integral:
\begin{equation}
S_{\mathrm{FW}}[x]
=
\frac{1}{4D}
\int_{t_i}^{t_f}
\left[
\dot{x}(t)-b(x,t)
\right]^2
\dd t.
\label{eq:app_fw_general}
\end{equation}
Equation~\eqref{eq:app_fw_general} is the Freidlin--Wentzell
rate-functional form appropriate to additive Gaussian noise
\cite{FreidlinWentzell2012}. With the present normalization, the
factor $1/D$ has been included in the definition of the dimensionless
stochastic action. Equivalently, one may define
\begin{equation}
I[x]
=
\frac{1}{4}
\int_{t_i}^{t_f}
\left[
\dot{x}(t)-b(x,t)
\right]^2
\dd t,
\end{equation}
so that the small-noise path probability has the asymptotic form
\begin{equation}
\mathcal P[x]
\approx 
\exp\left[-\frac{I[x]}{D}\right].
\label{eq:app_fw_asymptotic}
\end{equation}
The two conventions are equivalent, with
$S_{\mathrm{FW}}=I/D$.

Finally, substituting the gradient drift from
Equation~\eqref{eq:app_gradient_drift} into
Equation~\eqref{eq:app_fw_general} gives
\begin{align}
\dot{x}-b(x,t)
&=
\dot{x}
+
\frac{1}{\gamma}
\frac{\partial U_{\mathrm{eff}}}{\partial x},
\end{align}
and therefore
\begin{equation}
S_{\mathrm{micro}}[x]
=
\frac{1}{4D}
\int_{t_i}^{t_f}
\left[
\dot{x}
+
\frac{1}{\gamma}
\frac{\partial U_{\mathrm{eff}}}{\partial x}
\right]^2
\dd t.
\label{eq:app_micro_action}
\end{equation}
This is Equation~\eqref{eq:micro_action} of the main text.

The physical meaning of Equation~\eqref{eq:app_micro_action} follows
directly from its derivation. The quantity
\begin{equation}
\dot{x}-b(x,t)
=
\dot{x}
+
\frac{1}{\gamma}
\frac{\partial U_{\mathrm{eff}}}{\partial x}
\end{equation}
measures the instantaneous deviation of a trajectory from the
deterministic guidance dynamics. A path that closely follows
$\dot{x}=b(x,t)$ requires only small stochastic fluctuations and
therefore accumulates a small value of the rate functional.
Conversely, a trajectory that repeatedly moves against the
deterministic drift requires larger noise fluctuations and accumulates
a larger stochastic action. Thus, for specified initial and final
conditions, paths minimizing Equation~\eqref{eq:app_micro_action} are
the exponentially favored trajectories in the Freidlin--Wentzell
small-noise sense.

We note that it is important to distinguish this rate functional from the exact
Onsager--Machlup continuum path functional. For a nonlinear drift, an
Onsager--Machlup representation may contain an additional
drift-divergence or Jacobian contribution, schematically proportional
to $\partial b/\partial x$, depending on the definition of the
continuum path measure~\cite{OnsagerMachlup1953}. Such a term does not
alter the leading Freidlin--Wentzell large-deviation functional
derived above. Moreover, for direct analysis of experimentally
sampled growth cone trajectories, the discrete expression in
Equation~\eqref{eq:app_discrete_action} provides the most direct
likelihood-based definition of the stochastic path cost.

\section{Derivation of the Quasi-Static Beam Curvature Relation}
\label{app:beam_derivation}

Here we derive the quasi-static beam relations used in
Section~\ref{sec:durotaxis_action} to connect stiffness-dependent growth cone
traction to the curvature of the distal axon. The derivation follows the
small-deflection beam description used in mechanical models of axonal
durotaxis~\cite{Oliveri2021,Kassianides2025}.

Consider the mechanically active distal segment of the axon as an
Euler--Bernoulli beam of length $L$ and bending stiffness $\mathcal B$.
Let $z$ (with $0 \le z \le L$) denote distance along the undeformed axonal axis, with
$z=0$ corresponding to the proximal end of the active segment and $z=L$
to the growth cone. The transverse deflection is denoted by $y(z)$.
The growth cone exerts a tangential tensile force $F_t$ and a transverse
force $F_n$ on the distal segment. We assume small slopes,
$|y'(z)|\ll1$, and treat $F_t$ as approximately constant over the active
region during the rapid mechanical relaxation.

For a small transverse deformation, the elastic bending energy is
$\frac{1}{2}\mathcal B\int_0^L[y''(z)]^2\,dz$~\cite{Barber2022}. The axial tensile force
also contributes through the increase in contour length, which to
quadratic order in the slope is
$\frac{1}{2}\int_0^L[y'(z)]^2\,dz$. The corresponding potential energy
functional can therefore be written as
\begin{equation}
\Pi[y]
=
\frac{\mathcal B}{2}
\int_0^L [y''(z)]^2\,dz
+
\frac{F_t}{2}
\int_0^L [y'(z)]^2\,dz
-
F_n y(L).
\label{eq:app_beam_energy}
\end{equation}
Taking the first variation and integrating by parts gives
\begin{equation}
\begin{split}
\delta\Pi
={}&
\int_0^L
\left(
\mathcal B y''''-F_t y''
\right)\delta y\,dz
\\
&+
\left[
\mathcal B y''\delta y'
+
\left(
-\mathcal B y'''+F_t y'
\right)\delta y
\right]_0^L
-
F_n\delta y(L).
\end{split}
\label{eq:app_beam_variation}
\end{equation}

The proximal end of the active segment is taken to be clamped, so that
$y(0)=0$ and $y'(0)=0$. Because the corresponding variations vanish at
$z=0$, stationarity of Equation~\eqref{eq:app_beam_energy} for arbitrary
variations in the interior gives
\begin{equation}
\mathcal B y''''-F_t y''=0,
\label{eq:app_beam_static}
\end{equation}
which is Equation~\eqref{eq:beam_static} in the main text. The natural
boundary conditions at the growth cone are
\begin{equation}
y''(L)=0,
\qquad
\mathcal B y'''(L)-F_t y'(L)=-F_n.
\label{eq:app_beam_bc}
\end{equation}
The first condition expresses the absence of an externally imposed bending
moment at the growth cone, while the second represents transverse force
balance at the end of the beam.

To obtain the curvature at the base of the active segment, we define the local
slope $p(z)=y'(z)$. Since Equation~\eqref{eq:app_beam_static} can be
integrated once and the second boundary condition in
Equation~\eqref{eq:app_beam_bc} fixes the integration constant, $p(z)$
satisfies
\begin{equation}
\mathcal B p''-F_t p=-F_n.
\label{eq:app_slope_equation}
\end{equation}
Introducing
\begin{equation}
q=\sqrt{\frac{F_t}{\mathcal B}},
\label{eq:app_q}
\end{equation}
the general solution is
\begin{equation}
p(z)
=
\frac{F_n}{F_t}
+
C_1\cosh(qz)
+
C_2\sinh(qz).
\label{eq:app_slope_solution}
\end{equation}

The clamped condition $p(0)=0$ gives
$C_1=-F_n/F_t$. The zero-moment condition at the growth cone,
$p'(L)=y''(L)=0$, then gives
\begin{equation}
C_2
=
\frac{F_n}{F_t}
\tanh(qL).
\label{eq:app_C2}
\end{equation}

The curvature at the proximal end of the mechanically active region is,
within the small-slope approximation, $\kappa=y''(0)=p'(0)$. Using
Equations~\eqref{eq:app_q} and \eqref{eq:app_C2} therefore gives
\begin{equation}
\kappa
=
\frac{F_n}{\sqrt{\mathcal B F_t}}
\tanh\left(
L\sqrt{\frac{F_t}{\mathcal B}}
\right),
\label{eq:app_beam_curvature}
\end{equation}
which is Equation~\eqref{eq:beam_curvature} in the main text. The sign of
$\kappa$ follows the chosen sign convention for the transverse force $F_n$
and the trajectory normal. Equation~\eqref{eq:app_beam_curvature} therefore
also applies as a signed curvature relation.

As a consistency check, in the limit of vanishing tangential tension,
$F_t\rightarrow0$, the expansion $\tanh x\simeq x$ gives
\begin{equation}
\kappa
\longrightarrow
\frac{F_nL}{\mathcal B},
\label{eq:app_zero_tension}
\end{equation}
which is the standard base curvature of a cantilever beam subjected to a
transverse tip force~\cite{Barber2022}.

For the durotactic problem, the tangential and transverse forces are
determined by the stiffness-dependent growth cone traction. For an
approximately symmetric growth cone, the leading force components derived
from the finite growth cone model are~\cite{Oliveri2021,Kassianides2025}
\begin{equation}
F_t
\simeq
\alpha_g T(E_s),
\qquad
F_n
\simeq
\beta_g
\frac{dT}{dE_s}
\,\mathbf n\cdot\nabla E_s,
\label{eq:app_FtFn}
\end{equation}
where $E_s$ is the local substrate Young's modulus,
$T(E_s)$ is the corresponding traction, $\mathbf n$ is the local normal to
the axonal trajectory, $\alpha_g=2R\rho\sin\varphi_0$, and
$\beta_g=R^2\rho\varphi_0$.

Substitution of Equation~\eqref{eq:app_FtFn} into
Equation~\eqref{eq:app_beam_curvature} gives
\begin{equation}
\kappa
=
\frac{\beta_g}
{\sqrt{\mathcal B\alpha_g T}}
\tanh\left(
L\sqrt{\frac{\alpha_gT}{\mathcal B}}
\right)
\frac{dT}{dE_s}
\,\mathbf n\cdot\nabla E_s,
\label{eq:app_kappa_T}
\end{equation}
which is Equation~\eqref{eq:kappa_T} used in the main text. Since
$(dT/dE_s)\mathbf n\cdot\nabla E_s=\mathbf n\cdot\nabla T$, this relation
shows explicitly that the local axonal curvature is controlled by the
traction variation sampled in the direction normal to the trajectory.

The derivation assumes small axonal deflections, quasi-static mechanical
relaxation, an approximately constant tangential force over the distal active
segment, and a growth cone geometry for which the leading transverse response
can be represented by Equation~\eqref{eq:app_FtFn}. For the two-region PDMS
substrate considered in this work, the mathematically sharp stiffness boundary
should be understood as a finite mechanical transition region sampled by the
growth cone. Equation~\eqref{eq:app_kappa_T} describes the local turning
within this transition region, while the Fermat-like treatment in the main
text relates the incident and transmitted trajectories in the approximately
homogeneous regions on either side of the interface.

\section{Derivation of the Fermat-Like Curvature Relation}
\label{app:fermat_curvature}

Here we derive the curvature relation associated with the Fermat-like action
introduced in Equation~\eqref{eq:fermat_action}. Let the axonal trajectory be
represented by a planar curve $\mathbf r(u)$, where $u$ is an arbitrary path
parameter. The spatial action can then be written as
\begin{equation}
\mathcal A_{\mathrm F}[\mathbf r]
=
\int_{u_i}^{u_f}
n_{\mathrm{ax}}[\mathbf r(u)]
\left|\frac{d\mathbf r}{du}\right|du.
\label{eq:app_fermat_general}
\end{equation}
where $n_{\mathrm{ax}}(\mathbf r)$ is the effective axonal refractive index.
The integrand in Equation~\eqref{eq:app_fermat_general} therefore plays the
role of a spatial Lagrangian,
\begin{equation}
\mathcal L
=
n_{\mathrm{ax}}(\mathbf r)
|\mathbf r'|,
\qquad
\mathbf r'=\frac{d\mathbf r}{du}.
\label{eq:app_fermat_lagrangian}
\end{equation}

For fixed endpoints, stationarity of the action,
$\delta\mathcal A_{\mathrm F}=0$, gives the Euler--Lagrange equation
\begin{equation}
\frac{d}{du}
\left(
\frac{\partial\mathcal L}{\partial\mathbf r'}
\right)
-
\frac{\partial\mathcal L}{\partial\mathbf r}
=
0.
\label{eq:app_fermat_EL}
\end{equation}
Introducing the unit tangent vector
$\mathbf t=\mathbf r'/|\mathbf r'|$, the derivatives of the Fermat-like
Lagrangian with respect to $\mathbf r'$ and $\mathbf r$ are
\begin{equation}
\frac{\partial\mathcal L}{\partial\mathbf r'}
=
n_{\mathrm{ax}}\mathbf t,
\qquad
\frac{\partial\mathcal L}{\partial\mathbf r}
=
|\mathbf r'|\nabla n_{\mathrm{ax}}.
\label{eq:app_fermat_derivatives}
\end{equation}
Substitution into the Euler--Lagrange equation gives
\begin{equation}
\frac{d}{du}
\left(
n_{\mathrm{ax}}\mathbf t
\right)
=
|\mathbf r'|\nabla n_{\mathrm{ax}}.
\label{eq:app_ray_u}
\end{equation}
Since the arclength element satisfies
$ds=|\mathbf r'|\,du$, we have
$d/du=|\mathbf r'|\,d/ds$. Dividing
Equation~\eqref{eq:app_ray_u} by $|\mathbf r'|$ therefore gives
\begin{equation}
\frac{d}{ds}
\left(
n_{\mathrm{ax}}\mathbf t
\right)
=
\nabla n_{\mathrm{ax}}.
\label{eq:app_ray_equation}
\end{equation}

This is the spatial Euler--Lagrange equation corresponding to the Fermat-like
action~\cite{BornWolf1999}.

To obtain the curvature relation, expand the derivative on the left-hand side:
\begin{equation}
\frac{d}{ds}
\left(
n_{\mathrm{ax}}\mathbf t
\right)
=
\frac{dn_{\mathrm{ax}}}{ds}\mathbf t
+
n_{\mathrm{ax}}\frac{d\mathbf t}{ds}.
\label{eq:app_ray_expand}
\end{equation}

Along the trajectory,
\begin{equation}
\frac{dn_{\mathrm{ax}}}{ds}
=
\mathbf t\cdot\nabla n_{\mathrm{ax}},
\label{eq:app_index_derivative}
\end{equation}
while the Frenet relation for a planar curve is
\begin{equation}
\frac{d\mathbf t}{ds}
=
\kappa\mathbf n,
\label{eq:app_frenet}
\end{equation}
where $\mathbf n$ is the unit normal to the trajectory and $\kappa$ is the
signed curvature. Equation~\eqref{eq:app_ray_expand} consequently becomes
\begin{equation}
\left(
\mathbf t\cdot\nabla n_{\mathrm{ax}}
\right)\mathbf t
+
n_{\mathrm{ax}}\kappa\mathbf n
=
\nabla n_{\mathrm{ax}}.
\label{eq:app_ray_components}
\end{equation}
Taking the scalar product of Equation~\eqref{eq:app_ray_components} with the
unit normal $\mathbf n$, and using $\mathbf n\cdot\mathbf t=0$:
\begin{equation}
n_{\mathrm{ax}}\kappa
=
\mathbf n\cdot\nabla n_{\mathrm{ax}}.
\label{eq:app_normal_component}
\end{equation}
Therefore,
\begin{equation}
\kappa
=
\frac{\mathbf n\cdot\nabla n_{\mathrm{ax}}}
{n_{\mathrm{ax}}}
=
\mathbf n\cdot\nabla\ln n_{\mathrm{ax}}.
\label{eq:app_fermat_curvature}
\end{equation}

Equation~\eqref{eq:app_fermat_curvature} is the curvature relation used in the
main text. It shows that, within the Fermat-like approximation, an axonal
trajectory bends in response to the component of the effective-index gradient
normal to its instantaneous direction of growth. A spatial variation of
$n_{\mathrm{ax}}$ purely along the tangent changes the local value of the
index but does not directly generate curvature. Conversely, a transverse
gradient produces turning of the trajectory.

For the stiffness-dependent problem considered here,
$n_{\mathrm{ax}}=n_{\mathrm{ax}}[T(E_s)]$. The chain rule therefore gives
\begin{equation}
\nabla\ln n_{\mathrm{ax}}
=
\frac{d\ln n_{\mathrm{ax}}}{dT}
\frac{dT}{dE_s}
\nabla E_s,
\label{eq:app_index_chain}
\end{equation}
and Equation~\eqref{eq:app_fermat_curvature} becomes
\begin{equation}
\kappa
=
\frac{d\ln n_{\mathrm{ax}}}{dT}
\frac{dT}{dE_s}
\mathbf n\cdot\nabla E_s.
\label{eq:app_fermat_curvature_stiffness}
\end{equation}

This form provides the direct connection to the mechanical curvature relation
derived in Appendix~\ref{app:beam_derivation}. Comparing
Equation~\eqref{eq:app_fermat_curvature_stiffness} with the mechanically
derived expression for $\kappa$ determines the effective index relation
$d\ln n_{\mathrm{ax}}/dT$ used in Equation~\eqref{eq:index_diff}.

\section{{Experimental Data and Statistical Analysis}}
\label{app:fig4_analysis}

\begin{figure}[t]
    \includegraphics[width=0.65\textwidth]{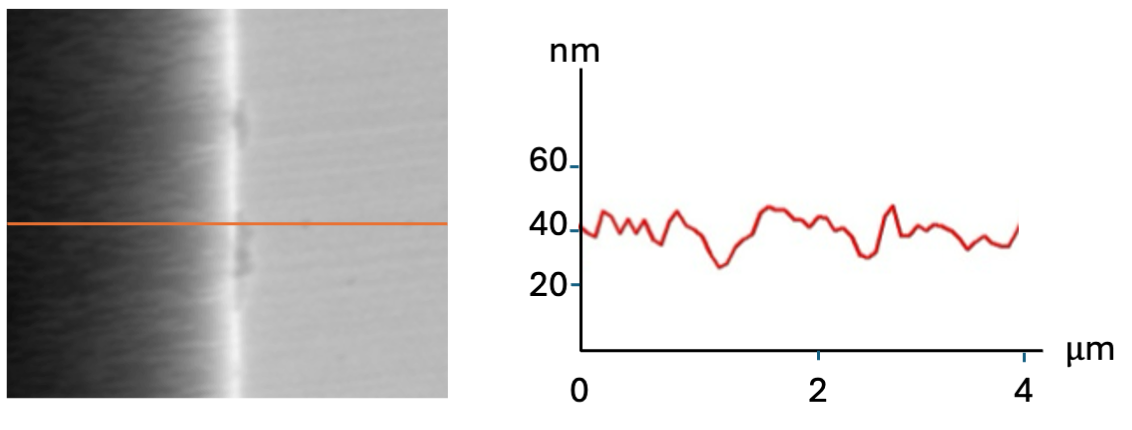}
  \caption{AFM characterization of the interface between the two PDMS regions of
different stiffness. \textbf{Left:} 
 representative $4\times4~\upmu\mathrm{m}^2$
topographic AFM image spanning the interface.  \textbf{Right:} corresponding line
profile across the interface, showing surface roughness on the order of
$20$~nm and a transition region approximately $0.5~\upmu$m in width. No significant topographical
discontinuity is observed across the boundary.}
     \label{figA1}
\end{figure}

The topography of the interface between the two PDMS regions was further
characterized by AFM. A representative $4\times4~\upmu\mathrm{m}^2$ AFM
topographic image and the corresponding line profile are shown in
Figure~\ref{figA1}. The surface roughness is on the order of
$20$~nm, with no significant topographical step across the interface. The
transition region has a characteristic width of approximately $0.5~\upmu$m,
indicating that the stiffness boundary is spatially localized while remaining
topographically smooth on the scale relevant to growth cone motion.

{Importantly, the height profile shows no systematic topographical step
coincident with the stiffness boundary. Thus, within the spatial and vertical
resolution of the AFM measurements, the interface does not present a ridge,
groove, or height discontinuity comparable to the topographical guidance
features used in the micropattern experiments. Because both regions also
received the same PDL coating after fabrication, the cell-contacting surface
was treated uniformly across the interface.}

The stochastic-action analyses shown in Figure~\ref{fig4} were performed using
experimental measurements of $N=1840$ growth cones obtained in our
previous studies of cortical neurons growing on PDMS substrates containing
parallel micropatterns
\cite{Basso2019,Yurchenko2021,Yurchenko2019,Vensi2019,
Descoteaux2022,Staii2023,Staii2025}.
{The 1840 growth cone trajectories correspond to 1840 distinct neurons
obtained from seven independent neuronal culture preparations derived from a
total of 48 embryos and grown on 12 separate micropatterned PDMS substrates.
Thus, the dataset contains 1840 neuron-level trajectory measurements distributed
across seven independent culture preparations and 12 experimental substrates.}
No new neuronal cultures were generated specifically for this analysis. The
experimental growth cone trajectories and the corresponding angular and
spatial statistics were therefore previously reported, whereas the
stochastic-action analysis presented in Figure~\ref{fig4} is performed here
for the present work. {Each trajectory from a distinct neuron was treated
as one observational unit for the action analysis, while the seven independent
culture preparations represent the biological replication of the dataset.}
The $20~\upmu$m axonal segments used to construct angular distributions were not
treated as independent growth cones in the statistical analysis of
Figure~\ref{fig4}.

\subsection{Stochastic Parameters Used in the Action Analysis}

The parameters entering the stochastic action were obtained from the
Langevin and Fokker--Planck analyses developed and calibrated in our previous
work
\cite{Yurchenko2019,Vensi2019,Descoteaux2022,Staii2023,Staii2025}.
For the transverse growth cone dynamics, the effective stochastic model is
specified by the motility coefficient $D$, damping coefficient $\gamma$,
substrate-potential strength $U_0$, and feedback-potential strength $U_F$.
These quantities determine the deterministic drift and stochastic
fluctuations entering Equation~\eqref{eq:discrete_action}.

For untreated cortical neurons growing on micropatterned PDMS substrates with
$d=4~\upmu$m, previous fits of the spatial Fokker--Planck model to the
experimental growth cone distributions gave
\cite{Descoteaux2022}
\begin{equation}
D=(22\pm4)~\upmu{\rm m}^2{\rm h}^{-1},
\qquad
\gamma=(0.14\pm0.05)~{\rm h}^{-1},
\label{eq:app_fig4_Dgamma}
\end{equation}

and
\begin{equation}
U_0=(1.9\pm0.3)~\upmu{\rm m}^2{\rm h}^{-2},
\qquad
U_F=(0.23\pm0.08)~{\rm h}^{-2}.
\label{eq:app_fig4_U}
\end{equation}

The action analysis was evaluated over parameter values representative of the
experimentally determined range for the micropatterned substrates. The ranges
used for evaluating the robustness of the Figure~\ref{fig4} analysis were:
$D=15$--$25~\upmu{\rm m}^2{\rm h}^{-1}$,
$\gamma=0.10$--$0.15~{\rm h}^{-1}$,
$U_0=0.5$--$2.0~\upmu{\rm m}^2{\rm h}^{-2}$, and
$U_F=0.05$--$0.25~{\rm h}^{-2}$.
These intervals are intended to represent the experimentally relevant
parameter range rather than confidence intervals for a single substrate
condition. In the previous measurements, the substrate and feedback
potentials were largest for micropattern periods producing the strongest
axonal alignment and decreased as the geometrical guidance became weaker
\cite{Descoteaux2022,Staii2023,Staii2025}.
\begin{table}[t]
\caption{{Stochastic 
 parameters used in the analysis of
Figure~\ref{fig4}. The $d=4~\upmu$m values are experimentally determined
values previously reported for untreated cortical neurons
\cite{Descoteaux2022}. The final column gives the representative parameter
range examined in the Figure~\ref{fig4} analysis.}}
\label{tab:fig4_parameters}
\centering
\footnotesize
\setlength{\tabcolsep}{3pt}
\renewcommand{\arraystretch}{1.15}

\begin{tabular}{llll}
\toprule
\textbf{Parameter} &
\textbf{Physical Meaning} &
\textbf{$d=4~\upmu$m} &
\textbf{Range Used in Analysis} \\
\midrule

$D$ &
Effective motility coefficient &
$22\pm4~\upmu{\rm m}^2{\rm h}^{-1}$ &
$15$--$25~\upmu{\rm m}^2{\rm h}^{-1}$ \\

$\gamma$ &
Effective damping coefficient &
$0.14\pm0.05~{\rm h}^{-1}$ &
$0.10$--$0.15~{\rm h}^{-1}$ \\

\midrule

$U_0$ &
Periodic substrate-potential strength &
$1.9\pm0.3~\upmu{\rm m}^2{\rm h}^{-2}$ &
$0.5$--$2.0~\upmu{\rm m}^2{\rm h}^{-2}$ \\

$U_F$ &
Feedback-potential strength &
$0.23\pm0.08~{\rm h}^{-2}$ &
$0.05$--$0.25~{\rm h}^{-2}$ \\

\bottomrule
\end{tabular}
\end{table}

For completeness, the exact parameter values associated with each
micropattern spatial period included in Figure~\ref{fig4} are summarized in
Table~\ref{tab:fig4_parameters_by_d}. These values were obtained from the
previously published Fokker--Planck analyses~\cite{Yurchenko2019,Vensi2019,Descoteaux2022,Staii2023,Staii2025} rather than introduced as new adjustable parameters in the present calculation of the stochastic action.
\begin{table}[t]
\caption{Experimental sample sizes $N_d$ and stochastic parameters for the
micropattern conditions included in Figure~\ref{fig4}.}
\label{tab:fig4_parameters_by_d}
\setlength{\tabcolsep}{3.8mm}
\begin{tabular}{cccccc}
\toprule
\boldmath{$d~(\upmu{\rm m})$} &
\boldmath{$N_d$} &
\boldmath{$D~(\upmu{\rm m}^2/{\rm h})$} &
\boldmath{$\gamma~({\rm h}^{-1})$} &
\boldmath{$U_0~(\upmu{\rm m}^2/{\rm h}^2)$} &
\boldmath{$U_F~({\rm h}^{-2})$} \\
\midrule
$\;1\;$ & $\;315\;$ & $\;16 \pm3\;$ & $\;0.11 \pm0.04\;$ &
$\;0.5\pm0.1\;$ & $\;0.08\pm0.04\;$ \\
$\;2\;$ & $\;378\;$ & $\;18 \pm3\;$ & $\;0.12 \pm0.03\;$ &
$\;0.7\pm0.1\;$ & $\;0.1\pm0.03\;$ \\
$4$ & $\;421\;$ & $22\pm4$ & $0.14\pm0.05$ &
$1.9\pm0.3$ & $0.23\pm0.08$ \\
$6$ & $\;407\;$ & $25\pm5$ & $0.15\pm0.05$ &
$1.8\pm0.2$ & $0.24\pm0.09$ \\
$10$ & $\;319\;$ & $20\pm5$ & $0.13\pm0.06$ &
$1.5\pm0.4$ & $0.15\pm0.09$ \\
\midrule
Total & $1840$ & & & & \\
\bottomrule
\end{tabular}
\end{table}

\subsection{Test of the Exponential Action--Probability Relation}

Figure~\ref{fig4}c tests the predicted exponential dependence of trajectory
probability on stochastic action. For the trajectory classes used in this
analysis, the empirical class frequency was calculated as
$P_k=N_k/N_{\mathrm{tot}}$, and the corresponding mean stochastic action was
denoted by $S_k$. The data were fitted by ordinary least squares to
\begin{equation}
\ln P_k=C+\beta S_k,
\label{eq:app_fig4_regression}
\end{equation}
where the stochastic-action prediction corresponds to $\beta=-1$. The
least-squares fit gave
\begin{equation}
\hat{\beta}=-0.98,
\qquad
{\rm SE}(\hat{\beta})=0.03.
\label{eq:app_fig4_slope}
\end{equation}
To test whether the measured slope is statistically consistent with the
predicted value, we used the null hypothesis
\begin{equation}
H_0:\beta=-1.
\end{equation}
The corresponding test statistic is
\begin{equation}
t=
\frac{\hat{\beta}-(-1)}
{{\rm SE}(\hat{\beta})}.
\label{eq:app_fig4_ttest}
\end{equation}
Using $\hat{\beta}=-0.98$ and
${\rm SE}(\hat{\beta})=0.03$ gives $t=0.67$. Thus, the fitted slope differs
from the predicted value of $-1$ by less than one standard error and is
statistically consistent with the stochastic action prediction
$\mathcal P[\Gamma]\propto\exp[-S(\Gamma)]$. Using the normal approximation,
the corresponding $95\%$ confidence interval for the fitted slope is
approximately $[-1.04,-0.92]$, which contains the predicted value $-1$.
It is important to distinguish the empirical probability of a trajectory
class from the probability density assigned to an individual path.
The quantity $P_k=N_k/N_{\mathrm{tot}}$ represents the observed frequency of
the $k$th trajectory class. Its comparison with the corresponding mean action
therefore constitutes a coarse grained experimental test of the exponential
action dependence rather than an identification of a class frequency with the
probability of one particular trajectory.
\begin{table}[t]
\caption{Statistical %: The two-part continued table was combined into a single table because
frequency and stochastic action in Figure~\ref{fig4}c.}
\label{tab:fig4_regression}
\centering
\small
\setlength{\tabcolsep}{6pt}
\renewcommand{\arraystretch}{1.15}

\begin{tabular}{lll}
\toprule
\textbf{Quantity} & \textbf{Value} \\
\midrule

Total number of growth cone trajectories &
$N_{\mathrm{tot}}=1840$ \\

Number of trajectory classes &
$K=10$ \\

Regression model &
$\ln P_k=C+\beta S_k$ \\

Predicted slope &
$-1$ \\

\midrule

Least-squares slope, $\hat{\beta}$ &
$-0.98$ \\

Standard error of slope &
$0.03$ \\

Tested null hypothesis &
$H_0:\beta=-1$ \\

Test statistic &
$t=0.67$ \\

Degrees of freedom &
$8$ \\

Two-sided $p$-value &
$0.52$ \\

Coefficient of determination &
$R^2=0.993$ \\

\bottomrule
\end{tabular}
\end{table}

\end{document}